\documentclass[12pt,notitlepage]{amsart}%
\usepackage{amssymb}
\usepackage{amsfonts}
\usepackage{graphicx}
\usepackage{amscd}
\usepackage{graphicx}
\usepackage{amsmath}
\usepackage{amsxtra}
\usepackage{footmisc}%
\theoremstyle{plain}

\newtheorem{notation}{Notation}

\newtheorem{remark}{Remark}

\numberwithin{equation}{section}
\numberwithin{theorem}{section}
\numberwithin{lemma}{section}
\numberwithin{proposition}{section}
\numberwithin{corollary}{section}

\ifx\pdfoutput\relax\let\pdfoutput=\undefined\fi
\newcount\msipdfoutput
\ifx\pdfoutput\undefined\else
\ifcase\pdfoutput\else
\ifx\paperwidth\undefined\else
\ifdim\paperheight=0pt\relax\else\pdfpageheight\paperheight\fi
\ifdim\paperwidth=0pt\relax\else\pdfpagewidth\paperwidth\fi
\fi\fi\fi
\begin{document}
\title[$2$-Adic QM and CTQWs on Graphs]{$2$-Adic quantum mechanics, continuous-time quantum walks, and the space discreteness}
\author[Z\'{u}\~{n}iga-Galindo]{W. A. Z\'{u}\~{n}iga-Galindo}
\address{University of Texas Rio Grande Valley\\
School of Mathematical \& Statistical Sciences\\
One West University Blvd\\
Brownsville, TX 78520, United States}
\email{wilson.zunigagalindo@utrgv.edu}

\begin{abstract}
We show that a large class of 2-adic Schr\"{o}dinger equations is the scaling
limit of certain continuous-time quantum Markov chains (CTQMCs). Practically,
a discretization of such an equation gives a CTQMC. As a practical result, we
construct new types of continuous time quantum walks (CTQWs) on graphs using
two symmetric matrices. One matrix describes the transport between nodes in
one direction, while the second describes the transport between nodes in the
opposite direction. This construction includes, as a particular case, the
CTQWs constructed using adjacency matrices. The final goal of this work is to
contribute to the understanding of the foundations of quantum mechanics (QM)
and the role of the hypothesis of the discreteness of space. The connection
between 2-adic QM and CTQWs shows that 2-adic QM has a physical meaning.
2-Adic QM is a nonlocal theory because the Hamiltonians used are nonlocal
operators, and consequently, spooky actions at a distance are allowed.
However, this theory is not a mathematical toy. The experimental confirmation
of the violation of Bell's inequality implies that this theory allows realism.
We pointed out several new research problems connected with the foundations of
quantum mechanics.

\end{abstract}
\maketitle

\section{Introduction}

Since the 1920's, Born, Einstein, Jordan, Pauli, among others, were convinced
that the classical space-time continuum must abandoned on the elementary
quantum scale. The reader may consult \cite{Capellmann} for an in-depth
historical review. For instance, according to \cite{Capellmann}, Einstein and
Born believed that the traditional concept of space-time of macroscopic
physics cannot simply be transferred to quantum physics. But all the efforts
(especially from Born and Jordan) to construct physical theories with discrete
space-times failed.

This work revolves around the foundations of QM, the hypothesis about the
discreteness of space, and quantum networks. We have divided the introduction
into several sections for the reader's convenience.

\subsection{Results}

In the 1930s, Bronstein showed that general relativity and quantum mechanics
imply that the uncertainty $\Delta x$ of any length measurement satisfies
$\Delta x\geq L_{\text{Planck}}$, where $L_{\text{Planck}}$ is the Planck
length, \cite{Bronstein}. A well-accepted interpretation of this inequality is
that there are no intervals, just points below the Planck length,
\cite{Varadarajan}-\cite{Volovich}. We argue that this last assertion has a
precise mathematical translation: at the Planck length, the space is a
completely disconnected space $\mathcal{X}$, which means that the only
non-trivial connected subsets are points. A connected subset is a mathematical
incarnation of an infinitely divisible object (a continuum). Thus, Bronstein
inequality drives naturally to study physical models on space-times of type
$\mathbb{R}\times\mathcal{X}$. In the 1980s, Volovich conjectured that
$\mathcal{X=}\mathbb{Q}_{p}^{3}$, where $\mathbb{Q}_{p}$ is the field of
$p$-adic numbers, \cite{Volovich}. If we use $\mathbb{R}\times\mathbb{Q}%
_{p}^{3}$ as a model of space-time, then, there are no continuous word lines
joining two different points in the space ($\mathbb{Q}_{p}^{3}$). The
assumption of the discreteness of space, which requires passing from
$\mathbb{R}\times\mathbb{R}^{3}$ to $\mathbb{R}\times\mathbb{Q}_{p}^{3}$ (or
$\mathbb{R}\times\mathcal{X}$), is incompatible with special and general relativity.

In the Dirac-von Neumann formulation of QM, the states of a closed quantum
system are vectors of an abstract complex Hilbert space $\mathcal{H}$, and the
observables correspond to linear self-adjoint operators in $\mathcal{H}$,
\cite{Dirac}-\cite{Takhtajan}. In $p$-adic QM, $\mathcal{H}=L^{2}%
(\mathbb{Q}_{p}^{N})$ or $\mathcal{H}=L^{2}(U)$, where $U\subset\mathbb{Q}%
_{p}^{N}$. This choice implies that the position vectors are $\mathbb{Q}%
_{p}^{N}$, while the time is a real variable. We warn the reader that there
are several different types of $p$-adic QM; for instance, if the time is
assumed to be a $p$-adic variable, the QM obtained radically differs from the
one considered here. To the best of our knowledge, $p$-adic QM started in the
1980s under the influence of Vladimirov and Volovich, \cite{V-V-QM3}. The
literature about $p$-adic QM is very extensive, and here we cited just a few
works, \cite{V-V-QM3}-\cite{Zuniga-PhA}.

This article continues our work on $p$-adic QM, \cite{Zuniga-AP}%
-\cite{Zuniga-PhA}.\ In \cite{Zuniga-AP}, a $p$-adic model of the double-slit
experiment was presented. The results of this work match the ones given in
\cite{Aharonov et al}: a localized particle with nonlocal interactions with
the other slit goes through one slit only. In \cite{Zuniga-PhA}, a new
$p$-adic Dirac equation that predicts the existence of pairs of
particle/antiparticle, and charge conjugation was introduced. These works
\ suggest that $p$-adic QM might describe physical systems. The main result of
this work is to confirm that $p$-adic QM describes physical systems, by
establishing a connection with continuous time quantum walks (CTQWs).

We show that a large class of $p$-adic Schr\"{o}dinger equations are
continuous versions (scaling limits) of certain continuous time quantum Markov
chains (CTQMCs). In practical terms, a discretization (in the sense of
numerical analysis) of a given a $p$-adic Schr\"{o}dinger equation gives a
CTQMC. These models include most of the continuous time quantum walks (CTQWs),
usually are formulated on graphs, as a particular case. The construction of
CTQMCs does not require using the adjacency matrix of a graph, but most of the
constructions of CTQWs, as far as we know, require adjacency matrices. We use
CTQMC instead of CTQW to emphasize that a model construction does not require
adjacency matrices.

CTQWs play a central in quantum transport and quantum computing,
\cite{Farhi-Gutman}-\cite{Zimboras et al}. As an application, we construct new
types of CTQWs on graphs using two symmetric matrices. One matrix describes
the transport between nodes in a one direction, while the second one describe
the transport between nodes in the opposite direction, see Section
\ref{Sect_6}. This construction includes, \ as a particular case, the
Farhi-Gutman CTQWs given in \cite{Farhi-Gutman}, and the ones given in
\cite{Mulkne-Blumen}- \cite{Venegas-Andraca}. The master equations of the
above-mentioned CTQWs are obtained as discretizations (in the sense of
numerical analysis) of $p$-adic Schr\"{o}dinger equations, see Section
\ref{Sect_6}.\ 

Given a \ $p$-adic Schr\"{o}dinger equation several different discretizations
are possible, each of them given a different CTQMC; see Sections \ref{Sect_5}
and \ref{Sect_6}. Here, two observations are required. First, there are CTQWs
obtained as discretizations of open quantum systems, see \cite{Frigeiro et
al.}, \cite{Zimboras et al}. This type of CTQW is out of this work's scope
because we work exclusively with closed quantum systems here. Second, here we
work with a rigorous mathematical notion of discretization, while in
\cite{Frigeiro et al.}, \cite{Zimboras et al} not. Later in the section on
open problems, we retake this discussion.

We constructed models of (closed) quantum systems on the space-time
$\mathbb{R}\times\mathbb{Q}_{p}$, which means that $\mathbb{Q}_{p}$, is the
correct model of space (or configuration space) for describing relevant
quantum phenomena, \ this suggests that the configuration spaces for such
phenomena are of the form $\mathbb{R}\times\mathcal{X}^{N}$, where
$\mathcal{X}$ is a completely disconnected space.

Then, the hypothesis about the discreteness of space at the Planck length
should be extended to general quantum phenomena, which is in line with the
historical discussion at the introduction's beginning. We warn the reader that
there are many publications on discrete QM, where the model of the space is a
lattice contained in $\mathbb{R}^{N}$. For $N=3$, this notion of discrete
space is compatible with special and general relativity and radically differs
from the one used in this work.

This work intends to contribute to the understanding of the foundations of QM,
and the role of the hypothesis of the discreteness of space as discussed in
the first lines of the introduction. The connection of $p$-adic QM and CTQWs
is probably profound fact, but in this work, the role of this connection is to
show that $p$-adic QM has a physical meaning. $p$-Adic QM is a nonlocal
theory, because the Hamiltonians used are nonlocal operators; consequently,
"spooky actions at a distance naturally occur." However, this theory is not a
mathematical toy because it describes a large class of quantum systems
(CTQWs). The experimental confirmation of the violation of Bell's inequality
leads to the paradigm that the universe is not locally real, which forces us
to pick between locality and realism, \cite{QN-1}-\cite{QN-5}. Since $p$-adic
QM is nonlocal, this theory admits a reality independent of the observer:
quantum particles exist, and they move randomly in a configuration space of
type $\mathbb{R}\times\mathcal{X}^{N}$, this means that the quantum motion can
be modeled as a Markov process in $\mathbb{R}\times\mathcal{X}^{N}%
$.\ Intuitively, the CTQWs introduced here are approximations of the quantum
motion mentioned.

On the other hand, in standard QM, the Hamiltonians are local operators; even
more, in standard quantum field theory, in order to apply functional
integration, the Lagrangian densities are required to be local in space and
time, \cite[p. 89]{Zinn-Justin}. In this framework, the consistency with the
mentioned paradigm represents a serious challenge.

\subsection{Open problems}

In our view, this work opens several new research areas. This section
discusses several open problems; others will be formulated later. For the sake
of simplicity, we assume that $\mathcal{X}=\mathbb{Q}_{p}$, but the discussion
can be easily extended to more general spaces. The success of the relativistic
QM, which uses $\mathbb{R}\times\mathbb{R}^{3}$as a model of the space-time,
is undeniable. On the other hand, we have shown that the space-time model
$\mathbb{R}\times\mathbb{Q}_{p}$ (or more generally $\mathbb{R}\times
\mathcal{X}^{N}$) is useful in formulation of quantum networks. A first
problem is how to unify these two models. Motivated by the work of Dragovich
and his collaborators, see e.g. \cite{Dragovich1}-\cite{Dragovich2}, it is
natural to propose the following model of space-time:%
\begin{equation}
\mathbb{R}\times\left(  \mathbb{R}\times\mathbb{Q}_{p}\right)  ^{3}.
\label{spacetime}%
\end{equation}
Using that $L^{2}\mathbb{\left(  R\right)  }$ and $L^{2}\left(  \mathbb{Q}%
_{p}\right)  $ are separable Hilbert spaces, the QM (in the sense of Dirac-von
Neumann) on the Hilbert space%
\[
\mathcal{H}=L^{2}\left(  \mathbb{R}\times\mathbb{Q}_{p}\right)  ^{3}%
\simeq\left(  L^{2}\left(  \mathbb{R}\right)
%TCIMACRO{\tbigotimes }%
%BeginExpansion
{\textstyle\bigotimes}
%EndExpansion
L^{2}\left(  \mathbb{Q}_{p}\right)  \right)  ^{3}%
\]
is well-defined. The space-time\ model (\ref{spacetime}) contains regions of
the form%
\[
\mathbb{R}\times\left(  \mathbb{R}\times\left\{  \alpha\right\}  \right)
^{3}\simeq\mathbb{R}\times\mathbb{R}^{3},
\]
for some fixed $\alpha\in\mathbb{Q}_{p}$, where the standard relativistic QM
can be formulated. The space model \ (\ref{spacetime}) also contains regions
of the form%
\[
\mathbb{R}\times\left(  \left\{  \beta\right\}  \times\mathbb{Q}_{p}\right)
^{3}\simeq\mathbb{R}\times\mathbb{Q}_{p}^{3},
\]
for some fixed $\beta\in\mathbb{R}$, where $p$-adic QM can be formulated.

A central question is whether this framework is useful in understanding the
wavefunction problem's collapse. The solution to the measurement problem is a
fundamental problem at the foundations of QM. A more doable problem consists
in the formulation of the Ghirardi, Ramini, and Weber (GRW) collapse models on
the space-time (\ref{spacetime}); see \cite{Norsen}-\cite{Ghiradi et al.}, and
the references therein.

The study of the GWR model involves the study of open quantum systems.
Studying $p$-adic quantum open systems is an open and relevant research
problem. Here, it is essential to mention that the theory of quantum Markovian
master equations is available on abstract Hilbert spaces, \cite{Gorinii et
al}, see also \cite{Lindblad}, in particular, for Hilbert spaces of type
$L^{2}\left(  U\right)  $, where $U$ is an open compact subset of
$\mathbb{Q}_{p}$. The discretization of $p$-adic versions of the mentioned
equations leads naturally to CTQWs as the ones considered in \cite{Frigeiro et
al.}, \cite{Zimboras et al}.

Along this work, we assume that the $p$-adic Schr\"{o}dinger equations are
obtained from $p$-adic heat equations by performing a Wick rotation. This
condition does not come from the Dirac-von Neumann axioms, but it is the
foundation of the path formulation of QM. A $p$-adic heat equation is an
evolution equation describing a random motion (a Markov process) in
$\mathbb{Q}_{p}^{N}$. For an introduction, the reader may consult
\cite{Kochubei}-\cite{Zuniga-Textbook}; the monograph \cite{Zuniga-LNM-2016}
exposes a general theory of such equations. A key fact is that the discrete
heat equation on a graph, which is constructed using the Laplacian of the
graph, is a $p$-adic heat equation. The author first reported this in
\cite{zuniga2020reaction}. The $p$-adic heat equations are obtained as
continuous limits of master equations of complex systems with a hierarchical
organization; see e.g., \cite{KKZuniga}-\cite{Zuniga-networks}, and the
references therein.

Here, we study the $p$-adic Schr\"{o}dinger equations obtained from the
$p$-adic heat equations considered in \cite{Zuniga-networks}; these equations
describe the dynamics of stochastic networks, and then, the equations obtained
by a Wick rotation described quantum versions of the mentioned networks. This
principle was also used in \cite{Farhi-Gutman}-\cite{Childs et al}.

A fundamental problem is to construct a mathematical theory for the CTQMCs
obtained by discretizing $p$-adic Schr\"{o}dinger equations.

\subsection{Further comments}

It is relevant to mention that our contribution fits in the proposal that, at
a fundamental level, the universe can be described as a cellular automaton or
a neural network; see, e.g., \cite{Vanchurin}-\cite{Hooft}. Before comparing
our contribution in relation to the mentioned literature, we say that the
$p$-adic analysis is a useful tool for understanding hierarchically organized
systems, see \cite{KKZuniga}, and the references therein. In particular, the
hierarchical structure of the $p$-adic numbers allows the construction of a
$p$-adic version of cellular neural networks that can be used in practical
tasks in image processing, see, e.g., \cite{Zuniga-Images}%
-\cite{Zuniga-Images-2}, and the references therein. $p$-Adic Euclidean
quantum field theories describe neural networks with hierarchical structures,
see, e.g., \cite{Zuniga-SFTS1}-\cite{Zuniga-SFTS3}, and the references therein.

`t Hooft has proposed that Nature at the Planck scale is an
information-processing machine like a deterministic cellular automaton,
\cite{Hooft}. He proposes a new QM in which the evolution of any quantum
system is entirely deterministic, and the probabilistic interpretation follows
from our inability to determine the initial state. In addition, his
deterministic QM is not consistent with the violation of Bell's inequality.
Here, we\ use stochastic automata, and the $p$-adic QM is consistent with the
violation of Bell's inequality. Finally, our results, provide a theoretical
support for the idea that the universe can be understood as a gigantic quantum
computer; see, e.g., \cite{Fredkin}-\cite{Zenil-Editor}.

\subsection{Road map}

The article is organized as follows. In Section \ref{Sect_1}, we introduce the
$p$-adic numbers and fix some notation. The Appendix summarizes the basic
results of $p$-adic analysis required here. For convenience, we take $p=2$,
but our results are valid for arbitrary $p$.

Section \ref{Sect_2} aims to provide a physical discussion about $p$-adic QM,
the Planck length, quantum nonlocality, the breaking of the Lorentz symmetry,
Einstein causality. The function spaces $L^{2}(\mathbb{R}^{N})$,
$L^{2}(\mathbb{Q}_{p}^{N})$ are isometric Hilbert spaces. Then, constructing
$p$-adic versions of standard quantum mechanical models is quite natural. For
instance, in \cite{Zuniga-PhA}, we introduced a $p$-adic Dirac equation, an
equation that shares many properties with the standard one. In particular, the
new equation also predicts the existence of pairs of particles and
antiparticles and a charge conjugation symmetry. In this framework, a
mathematical theorem shows that the Einstein causality is invalid in
$\mathbb{R}\times\mathbb{Q}_{p}^{3}$. Here, it is crucial to recall that
Eberhard and Ross, using $\mathbb{R}\times\mathbb{R}^{3}$ as a space-time
model, showed that the relativistic quantum field theory inherently forbids
faster-than-light communication, \cite{Eberhard et al}. This result is known
as the no-communication theorem. Therefore, whether the no-communication
theorem is valid if we replace $\mathbb{R}\times\mathbb{R}^{3}$ with
$\mathbb{R}\times\mathbb{Q}_{p}^{3}$ is unknown. This observation is relevant.
A well-spread belief is that the Eberhard and Ross theorem implies that
quantum nonlocality does not allow for faster-than-light communication and is
compatible with special relativity and its universal speed limit of objects.
This last assertion is true if quantum nonlocality is compatible with
\textquotedblleft\ $\mathbb{R}\times\mathbb{R}^{3}$ is a model of space-time
at all scales.\textquotedblright\ 

In section \ref{Sect_3}, we review the results of \cite{Zuniga-networks} about
$p$-adic heat equations and complex networks and introduce the Schr\"{o}dinger
equations required here. At the end of this section, \ we show how the Born
rule allows the construction of CTQWs on complete graphs, if an explicit
solution of the Cauchy problem attached to the Schr\"{o}dinger equation is
known, see Subsection \ref{Sub_CTQMC and Born Rule}. The drawback of this
approach is that explicit solutions of the Cauchy problem of the
Schr\"{o}dinger equation are required, which is generally a complicated task.
The approach given in this work avoids this problem.\ In \cite{Zuniga-Mayes},
a $p$-adic version of the infinite potential well in QM was studied. In this
case, it is possible to compute explicitly the solution of the Cauchy problem
for the Schr\"{o}dinger equation. In this framework, the CTWQs mentioned above
can be constructed.

Section \ref{Sect_4} is dedicated to the discretization of $2$-adic
Schr\"{o}dinger equations, also we review the construction of CTQWs following
\cite{Mulkne-Blumen}-\cite{Venegas-Andraca}. This article uses $2$-adic
Schr\"{o}dinger equations obtained by a Wick rotation from the $p$-adic heat
equations introduced in \cite{Zuniga-networks}. The choice $p=2$ simplifies
the comparison between some standard constructions in quantum computing with
our results. We show the existence of discretizations of the mentioned
equations of form $i\frac{\partial}{\partial t}\Psi_{I}(t)=\boldsymbol{H}%
^{\left(  r\right)  }\Psi_{I}(t)$, $I\in G_{l}^{0}$, where $G_{r}^{0}$ is a
finite set, with cardinality $\#G_{r}^{0}$, where $\Psi_{I}(t)\in
\mathbb{C}^{\#G_{r}^{0}}$ and $\boldsymbol{H}^{\left(  r\right)  }$ is a
matrix of size $\#G_{r}^{0}\times\#G_{r}^{0}$. Furthermore, $\mathbb{C}%
^{\#G_{r}^{0}}$ is isometric to a finite-dimensional subspace of test
functions $\mathcal{X}_{r}(\mathbb{Z}_{2})$, \ where $\mathbb{Z}_{2}$ is the
unit ball in $\mathbb{Q}_{2}$. At the end of this section, we use the matrices
$\boldsymbol{H}^{\left(  r\right)  }$ to construct CTQMCs (quantum networks),
see Subsection \ref{Sub_Quantum_networks}.

In Section \ref{Sect_5}, we show that the CTQWs on finite graphs, introduced
by Farhi and Gutmann, \cite{Farhi-Gutman}-\cite{Childs et al}, are particular
cases of the CTQMCs introduced here. At the end of this section, we consider
the existence of scaling limit for some $2$-adic Schr\"{o}dinger equations,
see Subsection \ref{Subsection _Scaling_Limit}. The solutions of these
discrete equations produce good approximations of the solution of the original
continuous equation. Starting with one of these discrete equations, we show
the existence of a continuous version of it as a $2$-adic Schr\"{o}dinger
equation $i\frac{\partial}{\partial t}\Psi(x,t)=\boldsymbol{H}\Psi(x,t)$,
where \ $\Psi(\cdot,t)\in L^{2}(\mathbb{Z}_{2})$, and $\boldsymbol{H}$ is a
self-adjoint operator on $L^{2}(\mathbb{Z}_{2})$.

Section \ref{Sect_6} introduces a new class of CTQWs on oriented graphs that
generalizes the construction of CTQWs based on adjacency matrices.

\section{\label{Sect_1}Analysis on the ring of $2$-adic numbers}

Any non-zero $2-$adic integer $x$ has a unique expansion of the form%
\begin{equation}
x=2^{\gamma}%
%TCIMACRO{\dsum \limits_{k=0}^{\infty}}%
%BeginExpansion
{\displaystyle\sum\limits_{k=0}^{\infty}}
%EndExpansion
x_{k}2^{k},\text{ }\gamma\in\mathbb{N}\text{, }x_{k}\in\left\{  0,1\right\}
\text{, and }x_{0}\neq0. \label{p-adic-integer}%
\end{equation}
The set of all possible sequences of the form (\ref{p-adic-integer})
constitutes the ring of $2$-adic integers $\mathbb{Z}_{2}$. There are natural
operations, sum and multiplication, on series of form (\ref{p-adic-integer}).
There is also a norm in $\mathbb{Z}_{2}$ defined as $\left\vert x\right\vert
_{2}=2^{-ord(x)}$, where $ord(x)=\gamma$, for a nonzero $2$-adic integer $x$.
By definition $ord(0)=\infty$. The ring $\mathbb{Z}_{2}$ with the distance
induced by $\left\vert \cdot\right\vert _{2}$ is a complete ultrametric space.
The ultrametric property refers to the fact that $\left\vert x-y\right\vert
_{2}\leq\max\left\{  \left\vert x-z\right\vert _{2},\left\vert z-y\right\vert
_{2}\right\}  $ for any $x$, $y$, $z$ in $\mathbb{Z}_{2}$.

The ball $B_{-l}(a)$ with center at $a\in\mathbb{Q}_{2}$ and radius $2^{-r}$,
$r\in\mathbb{N}$, is the set%
\[
B_{-r}(a)=a+2^{r}\mathbb{Z}_{2}=\left\{  x\in\mathbb{Z}_{2};\left\vert
x-a\right\vert _{2}\leq2^{-r}\right\}  \text{.}%
\]
Notice that $\mathbb{Z}_{2}=B_{0}(0)$ is the unit ball in $\mathbb{Q}_{2}$.
All the balls are infinite rooted trees with fractal structure.

A function $\varphi:\mathbb{Z}_{2}\rightarrow\mathbb{C}$ is called test
function, if there is a non-negative integer $l=l\left(  \varphi\right)  $
such that for any $a\in\mathbb{Z}_{2}$,
\[
\varphi\left(  a+x\right)  =\varphi\left(  a\right)  \text{ for any }%
|x|_{2}\leq2^{-l}.
\]
We denote by $\mathcal{D}(\mathbb{Z}_{2})$ the complex vector space of test
functions on $\mathbb{Z}_{2}$. There is a natural integration theory so that
$\int_{\mathbb{Z}_{2}}\varphi\left(  x\right)  dx$ gives a well-defined
complex number. The measure $dx$ is the Haar measure of $\mathbb{Z}_{2}$.

The Hilbert space of complex-valued, square-integrable functions defined on
$\mathbb{Z}_{2}$ is%
\[
L^{2}(\mathbb{Z}_{2}):=L^{2}(\mathbb{Z}_{2},dx)=\left\{  f:\mathbb{Z}%
_{2}\rightarrow\mathbb{C};\left\Vert f\right\Vert _{2}=\sqrt{\left\langle
f,f\right\rangle }<\infty\right\}  ,
\]
where%
\[
\left\langle f,g\right\rangle =%
%TCIMACRO{\dint \limits_{\mathbb{Z}_{2}}}%
%BeginExpansion
{\displaystyle\int\limits_{\mathbb{Z}_{2}}}
%EndExpansion
f\left(  x\right)  \overline{g\left(  x\right)  }dx\text{, \ }f,g\in
L^{2}(\mathbb{Z}_{2}),
\]
is the inner product; here the bar denotes the complex conjugate.

In the Appendix (Section \ref{Appendix}), we give a quick review of the basic
aspects of the $2$-adic analysis required here. For an in-depth exposition,
the reader may consult \cite{V-V-Z}, \cite{Kochubei}-\cite{Zuniga-Textbook},
\cite{Taibleson}-\cite{Alberio et al},

\begin{remark}
Let $p$ be a prime number. Any non-zero $p$-adic number $x$ has a unique
expansion of the form%
\begin{equation}
x=x_{-k}p^{-k}+x_{-k+1}p^{-k+1}+\ldots+x_{0}+x_{1}p+\ldots,\text{ }
\label{p-adic-number}%
\end{equation}
with $x_{-k}\neq0$, where $k$ is an integer, and the $x_{j}$s \ are numbers
from the set $\left\{  0,1,\ldots,p-1\right\}  $. The set of all possible
sequences form the (\ref{p-adic-number}) constitutes the field of $p$-adic
numbers $\mathbb{Q}_{p}$. There are natural field operations, sum and
multiplication, on series of form (\ref{p-adic-number}). There is also a norm
in $\mathbb{Q}_{p}$ defined as $\left\vert x\right\vert _{p}=p^{k}$, for a
nonzero $p$-adic number $x$. Along this article we work with $p=2$ for the
sake of simplicity, but the results are valid for arbitrary $p$. The main
reason to take $p=2$ is that the discretization of some $2$-adic
Schr\"{o}dinger operators are Hermitian matrices, which are associated with
continuous time quantum walks (CTQWs) on the Hilbert space $\mathbb{C}^{N}$,
$N=2^{l}$, that includes, as a particular case, Farhi-Gutmann CTQWs, see
\cite{Farhi-Gutman}.
\end{remark}

\section{\label{Sect_2}State of the art and motivations}

This section aims to provide a picture of the connection of $p$-adic QM, with
quantum nonlocatity, and the Lorentz symmetry breaking at the Planck scale.

\subsection{The discreteness of space}

In the 1930s, Bronstein showed that general relativity and quantum mechanics
imply that the uncertainty $\Delta x$ of any length measurement satisfies
\begin{equation}
\Delta x\geq L_{\text{Planck}}:=\sqrt{\frac{\hbar G}{c^{3}}},
\label{Inequality}%
\end{equation}
where $L_{\text{Planck}}$ is the Planck length ($L_{\text{Planck}}%
\approx10^{-33}$ $cm$), \cite{Bronstein}. This inequality establishes an
absolute limitation on length measurements, so the Planck length is the
smallest possible distance that can, in principle, be measured. The standard
interpretation of Bronstein's inequality is that below the Planck length there
are no intervals just points. This interpretation has a precise mathematical
translation: below the Planck length, the space is\ a totally (or completely)
disconnected topological space, which is the non-trivial connected subsets are points.

The choice of $\mathbb{R}$ as a model of the unidimensional space is not
compatible with inequality (\ref{Inequality}) because $\mathbb{R}$ contains
intervals with arbitrarily small length. On the other hand, there are no
intervals in $\mathbb{Q}_{p}$, i.e., the non-trivial connected subsets are
points. So $\mathbb{Q}_{p}$ is the prototype of a `discrete space' with a very
rich mathematical structure. In the 80s, Volovich conjectured on the $p$-adic
nature of the space at the Planck scale, \cite{Volovich}.

We denote by $T_{j_{0}}$ the set of $2$-adic integers of the form $j_{0}%
+j_{1}2$, with $j_{0},j_{1}\in\left\{  0,1\right\}  $. This set is a rooted
tree, with root $j_{0}$, and one layer. The vertices on this layer correspond
with the points of the form $j_{0}+j_{1}p$. $\mathbb{Z}_{2}$ is the
self-similar set obtained \ as a tilling using $T_{0}$.\ Alternatively,
$\mathbb{Z}_{2}$ is the self-similar constructed from $T_{0}$ using the scale
transformations%
\[
x\rightarrow a+2^{L}x\text{, with }a\in\mathbb{Z}_{2}\text{, }L\in\mathbb{N}.
\]
Then, up to a scale transformation, the smallest distance\ between two
different points in $\mathbb{Z}_{2}$ is
\[
2^{-1}=\left\vert 0-2\right\vert _{2}.
\]
This means, that if we pick $\mathbb{Z}_{2}$ as model of the space, the Planck
length is normalized to be $2^{-1}$. If we change $\mathbb{Z}_{2}$\ by
$\mathbb{Q}_{p}$, the Planck length changes to $p^{-1}$.

\subsection{$p$-Adic QM}

In the Dirac-Von Neumann formulation of QM, \cite{Dirac}-\cite{von Neumann},
to every isolated quantum system there is associated a separable complex
Hilbert space $\mathcal{H}$ called the space of states. The states of a
quantum system are described by non-zero vectors from $\mathcal{H}$. Each
observable corresponds to a unique linear self-adjoint operator in
$\mathcal{H}$. The most important observable of a quantum system is its
energy. We denote the corresponding operator by $\boldsymbol{H}$. Let
$\Psi_{0}\in\mathcal{H}$ be the state at time $t=0$ of a certain quantum
system. Then, at time $t$, the system is represented by the vector
$\Psi\left(  t\right)  =\boldsymbol{U}_{t}\Psi_{0}$, where
\[
\boldsymbol{U}_{t}=e^{-it\boldsymbol{H}}\text{, }t\geq0,
\]
is a unitary operator called the evolution operator. It is essential that $t$
be a real variable. The vector function $\Psi\left(  t\right)  $ is the
solution Schr\"{o}dinger equation
\[
i\frac{\partial}{\partial t}\Psi\left(  t\right)  =\boldsymbol{H}\Psi\left(
t\right)  \text{, }%
\]
where $i=\sqrt{-1}$ and the Planck constant is assumed to be one. For an
in-depth discussion the reader may consult \cite{Berezin et al}%
-\cite{Takhtajan}.

For the sake of simplicity, we consider only one-dimensional systems. In
standard QM, the states of quantum systems are functions from spaces of type
$L^{2}(\mathbb{R})$. In this case, the wavefunctions have the form
$\Psi\left(  x,t\right)  :\mathbb{R}\times\mathbb{R}_{+}\rightarrow\mathbb{C}%
$, where $x\in\mathbb{R}$, $t\in\mathbb{R}_{+}=\left\{  t\in\mathbb{R}%
;t\geq0\right\}  $. This choice implies that the space is continuous, i.e.,
given two different points $x_{0}$, $x_{1}\in\mathbb{R}$ there exists a
continuous curve $X\left(  t\right)  :\left[  a,b\right]  \rightarrow
\mathbb{R}$ such that $X\left(  a\right)  =x_{0}$, $X\left(  b\right)  =x_{1}%
$. The Dirac-von Neumann formulation of QM does not rule out the possibility
of choosing a `discrete space,' i.e., we can take $\mathcal{H=}L^{2}%
(\mathbb{Q}_{p})$; in this case the wavefunctions have the from $\Psi\left(
x,t\right)  :\mathbb{Q}_{p}\times\mathbb{R}_{+}\rightarrow\mathbb{C}$, where
$x\in\mathbb{Q}_{p}$, $t\in\mathbb{R}_{+}$. The space $\mathbb{Q}_{p}$ is a
completely disconnected topological space, which is a topological space where
the connected components are the points and the empty set. In such a space, a
continuous curve joining two different points does not exist. This implies
that the word line notion, which is a fundamental pillar in the formulation of
special and general relativity, does not exist if we assume as a model of
physical space a totally disconnected space. Consequently, the $p$-adic QM is
incompatible with the special and general relativity. $p$-Adic QM is a model
of the standard QM at the Planck length, and its study is potentially useful
in the understanding of the unification QM and gravity, \cite{Rovelli et al},
\cite{Amelino-Camelia}. The testability theories like $\ p$-adic QM, string
theory and quantum gravity, that work at the Planck scale require accessing
incredibly high energy levels. So, the physical content of such theories is in
question. This article aims to show that certain $2$-adic Schr\"{o}dinger
equations describe continuous versions of quantum networks, which include the
CTQWs defined on oriented graphs as a particular case.

\subsection{Quantum nonlocality}

We select an evolution operator $e^{\tau\boldsymbol{H}_{0}}$, $\tau\geq0$,\ so
that it is a Feller semigroup. Then $u(x,t)=e^{\tau\boldsymbol{H}_{0}}%
u_{0}(x)$ is the solution of the evolution equation (a $p$-adic heat equation)
of the form
\begin{equation}
\frac{\partial}{\partial\tau}u\left(  x,\tau\right)  =\boldsymbol{H}%
_{0}u\left(  x,\tau\right)  \text{, }x\in\mathbb{Q}_{p}\text{, }\tau\geq0,
\label{Eq_1A}%
\end{equation}
with initial datum $u\left(  x,0\right)  =u_{0}\left(  x\right)  .$ The Feller
condition implies the existence of Markov process in $\mathbb{Q}_{p}$, with
discontinuous paths, attached to equation (\ref{Eq_1A}); see e.g.
\cite{V-V-Z}, \cite{Kochubei}, \cite{Zuniga-Textbook}, \cite{Zuniga-LNM-2016}.
Here, it is relevant to mention that all known operators $\boldsymbol{H}_{0}$
appearing in the $p$-adic heat equation are non-local. We now apply the Wick
rotation $\tau=it$, $t\geq0$, with $i=\sqrt{-1}$, and $\Psi\left(  x,t\right)
=u\left(  x,it\right)  $, to (\ref{Eq_1A}) to obtain the free, $p$-adic
Schr\"{o}dinger equation:%
\[
i\frac{\partial}{\partial t}\Psi\left(  x,t\right)  =-\boldsymbol{H}_{0}%
\Psi\left(  x,t\right)  \text{, }x\in\mathbb{Q}_{p}\text{, }t\geq0.
\]
The simplest choice for $\boldsymbol{H}_{0}$\ is $\boldsymbol{D}^{\alpha}$,
$\alpha>0$, the Taibleson-Vladimirov fractional, which is defined as%
\[
\boldsymbol{D}^{\alpha}\varphi\left(  x\right)  =\frac{1-p^{\alpha}%
}{1-p^{-\alpha-1}}%
%TCIMACRO{\dint \limits_{\mathbb{Q}_{p}}}%
%BeginExpansion
{\displaystyle\int\limits_{\mathbb{Q}_{p}}}
%EndExpansion
\frac{\varphi(z)-\varphi(x)}{|z-x|_{p}^{\alpha+1}}\,dz,
\]
for $\varphi$ a locally constant function with compact support. To see the
non-local nature of this operator, we take $\varphi\left(  x\right)  =1$ if
$|x|_{p}\leq1$, otherwise $\varphi\left(  x\right)  =0$, then%
\begin{align*}
\boldsymbol{D}^{\alpha}\varphi\left(  x\right)   &  =\frac{1-p^{\alpha}%
}{1-p^{-\alpha-1}}\left\{
%TCIMACRO{\dint \limits_{|z|_{p}\leq1}}%
%BeginExpansion
{\displaystyle\int\limits_{|z|_{p}\leq1}}
%EndExpansion
\frac{\varphi(z)-\varphi(x)}{|z-x|_{p}^{\alpha+1}}\,dz+%
%TCIMACRO{\dint \limits_{|z|_{p}>1}}%
%BeginExpansion
{\displaystyle\int\limits_{|z|_{p}>1}}
%EndExpansion
\frac{\varphi(z)-\varphi(x)}{|z-x|_{p}^{\alpha+1}}\,dz\right\} \\
&  =\left\{
\begin{array}
[c]{lll}%
-\frac{1-p^{\alpha}}{1-p^{-\alpha-1}}\left(  \text{ }%
%TCIMACRO{\dint \limits_{|z|_{p}>1}}%
%BeginExpansion
{\displaystyle\int\limits_{|z|_{p}>1}}
%EndExpansion
\frac{dz}{|z|_{p}^{\alpha+1}}\,\right)  & \text{if} & |x|_{p}\leq1\\
&  & \\
\frac{1-p^{\alpha}}{1-p^{-\alpha-1}}\frac{1}{|x|_{p}^{\alpha+1}} & \text{if} &
|x|_{p}>1.
\end{array}
\right.
\end{align*}

In QM, the fact that the Hilbert space of a composite system is the Hilbert
space tensor product of the state spaces associated with the component
systems, and the superposition principle allows the existence of entangled
states between two particles, which drives to the quantum nonlocality. Quantum
nonlocality has been experimentally verified under a variety of physical
assumptions, \cite{QN-1}-\cite{QN-5}, see also \cite{Norsen}.

By definition, $p$-adic QM is a nonlocal theory. Hence, the violation of
Bell's inequality (i.e., the paradigm: the universe is not locally real) does
not cause any trouble in $p$-adic QM. Now, for standard QM, the mentioned
paradigm causes serious trouble since standard QM is supposed to be a local
theory, and abandoning the idea that objects have definite properties
independent of observation seems to have profound epistemological consequences.

\subsection{Breaking of the Lorentz symmetry and the violation of Einstein
causality}

Taking $\mathbb{R}\times\mathbb{Q}_{p}^{3}$ as a space-time model, in $p$-adic
QM, the Lorentz symmetry is broken, since the time and position at not
interchangeable. The Lorentz symmetry is one of the most essential symmetries
of the quantum field theory. While the validity of this symmetry continues to
be verified with a high degree of precision \cite{Kostelecky-Russell}, in the
last thirty-five years, the experimental and theoretical studies of the
Lorentz breaking symmetry have been an area of intense research, see, e.g.,
the reviews \cite{Mariz et al}, \cite{Amelino-Camelia} and the references therein.

In \cite{Zuniga-PhA}, we introduced a $p$-adic Dirac equation that shares many
properties with the standard one. In particular, the new equation also
predicts the existence of pairs of particles and antiparticles and a charge
conjugation symmetry. The $p$-adic Dirac spinors depend on the standard
Pauli-Dirac matrices. The new equation is a version of the standard Dirac
equation at the Planck scale, where the breaking of the Lorentz symmetry
naturally occurs. In this framework, the Einstein causality is invalid valid
in $\mathbb{R}\times\mathbb{Q}_{p}^{3}$. The derivation of the $p$-adic Dirac
equation is based on the fact that the plane wave solutions of the standard
Dirac equation have natural $p$-adic analogs when one considers the position
and momenta as elements from $\mathbb{Q}_{p}^{3}$; the construction of these
analogs does not require Lorentz invariance, just a version of the
relativistic energy formula, with $c=1$. This last normalization, or a similar
one, is essential because, in the new theory, the speed of light is not the
upper bound for the speed at which conventional matter or energy can travel
through space.

The $p$-adic Dirac equation has the form%
\begin{equation}
i\frac{\partial}{\partial t}\Psi\left(  t,\boldsymbol{x}\right)  =\left(
\boldsymbol{\alpha}\cdot\nabla_{\wp}+\beta m\right)  \Psi\left(
t,\boldsymbol{x}\right)  ,\text{ }t\in\mathbb{R}\text{, }\boldsymbol{x=}%
\left(  x_{1},x_{2},x_{3}\right)  \in\mathbb{Q}_{p}^{3}, \label{Dirac_1}%
\end{equation}
where $\boldsymbol{\alpha}$, $\cdot$, $m$ have the standard meaning,
\[
\Psi^{T}\left(  t,\boldsymbol{x}\right)  =\left[
\begin{array}
[c]{cccc}%
\Psi_{1}\left(  t,\boldsymbol{x}\right)  & \Psi_{2}\left(  t,\boldsymbol{x}%
\right)  & \Psi_{3}\left(  t,\boldsymbol{x}\right)  & \Psi_{4}\left(
t,\boldsymbol{x}\right)
\end{array}
\right]  \in\mathbb{C}^{4},
\]
$\nabla_{\wp}^{T}=\left[
\begin{array}
[c]{ccc}%
\boldsymbol{D}_{x_{1}} & \boldsymbol{D}_{x_{2}} & \boldsymbol{D}_{x_{3}}%
\end{array}
\right]  $,\ where $\boldsymbol{D}_{x_{k}}$ denotes the Taibleson-Vladimirov
fractional derivative.

The geometry of space $\mathbb{Q}_{p}^{3}$ imposes substantial restrictions on
the solutions of (\ref{Dirac_1}). The $p$-adic Dirac equation admits
space-localized planes waves $\Psi_{\boldsymbol{rnj}}\left(  t,\boldsymbol{x}%
\right)  $ for any time $t\geq0$, which is, $\mathrm{supp}$ $\Psi
_{\boldsymbol{rnj}}\left(  t,\boldsymbol{\cdot}\right)  $ is contained in a
compact subset of $\mathbb{Q}_{p}^{3}$. This phenomenon does not occur in the
standard case; see, e.g., \cite[Section 1.8, Corollary 1.7]{Thaller}. On the
other hand, we compute the transition probability from a localized state at
time $t=0$ to another localized state at $t>0$, assuming that the space
supports of the states are arbitrarily far away. It turns out that this
transition probability is greater than zero for any time $t\in\left(
0,\epsilon\right)  $, for arbitrarily small $\epsilon$; see \cite[Theorem
9.1]{Zuniga-PhA}. Since this probability is nonzero for some arbitrarily small
$t$, the system has a nonzero probability of getting between the mentioned
localized states arbitrarily shortly, thereby propagating with superluminal
speed in $\mathbb{R}\times\mathbb{Q}_{p}^{3}$.

\subsection{Quantum nonlocality and faster-than-light communication}

In 1988, Eberhard and Ross, using $\mathbb{R}\times\mathbb{R}^{3}$ as a
space-time model, showed that the relativistic quantum field theory inherently
forbids faster-than-light communication, \cite{Eberhard et al}. This result is
known as the no-communication theorem. It preserves the principle of causality
in quantum mechanics and ensures that information transfer does not violate
special relativity by exceeding the speed of light. So, if the space is not
discrete at the Planck length, then faster-than-light communication is
impossible. Therefore, the no-communication theorem does not rule out the
possible superluminal speed in $\mathbb{R}\times\mathbb{Q}_{p}^{3}$. Indeed,
the study of the possible superluminal speed under the hypothesis that space
is completely disconnected at the Planck length is an open problem.

In \cite{Ghirardi et al}, the authors assert that quantum nonlocality does not
allow for faster-than-light communication and hence is compatible with special
relativity and its universal speed limit of objects. In this work, the authors
assume that \textquotedblleft... quantum mechanics is unrestrictedly valid,
i.e. if it governs both the evolutions of the system and of measuring
apparata, as well as their interactions.\textquotedblright\ This hypothesis is
very questionable since the measurement problem in QM is still open. We
believe that \cite{Ghirardi et al} asserts that if the measurement problem in
QM on $L^{2}(\mathbb{R}^{N})$ has a satisfactory solution, then
faster-than-light communication is ruled out.

The propagating with superluminal speed predicted in $p$-adic QM is aligned
with the broken Lorentz symmetry expected at the Planck length,
\cite{Amelino-Camelia}. This conjecture cannot be easily discarded since this
article shows that $p$-adic QM has a physical meaning. On the other hand,
\ the no-communication theorem under the discreteness of the space is a very
subtle and complicated problem. Assuming that space discreteness implies the
violation of Einstein's causality, it does not follow the possibility that one
observer can transmit information to another observer exceeding the speed of
light. The construction of a device that exploits the violation of the
Einstein causality is an entirely different problem from the existence of
\ the violation phenomena. We think this situation is similar to the relation
between general relativity and the existence of a time machine. General
relativity predicts that gravity can cause time to stretch or dilate, meaning
time passes slower in stronger gravitational fields. This fact does imply the
existence of a time machine.

\section{\label{Sect_3}$2$-Adic Schr\"{o}dinger equations and quantum
networks}

From now on, we take $p=2$. By $2$-adic QM, we mean QM in the sense of the
Dirac-von Neumann\ on the Hilbert space $L^{2}(\mathbb{Z}_{2})$. In $2$-adic
QM, the Schr\"{o}dinger equations are obtained from $2$-adic heat equations by
performing a Wick rotation. These last equations are associated with Markov
processes, which are generalizations of the random motion of a particle in a
fractal, such as $\mathbb{Z}_{2}$ or $\mathbb{Q}_{2}$; see, e.g.,
\cite{Zuniga-Textbook}, \cite{Zuniga-LNM-2016}, \cite{KKZuniga}, and the
references therein. In \cite{Zuniga-networks}, see also \cite{Zuniga-Anselmo},
\cite{zuniga2020reaction}, we introduce a new type of stochastic networks,
which are $p$-adic continuous analogs of the standard Markov state models
constructed using master equations. The mentioned network is a model of a
complex system whose dynamics is described by a Markov process controlled
master equation. This master equation is a $2$-adic heat equation; by
performing a Wick rotation, we obtained a Schr\"{o}dinger equation that
describes a quantum version of the networks studied in \cite{Zuniga-networks}.

\subsection{$2$-Adic Heat equations and ultrametric networks}

We review some results of \cite{Zuniga-networks} required here. Let
$\mathcal{K}\subset\mathbb{Z}_{2}$ be a compact open subset, i.e.,
$\mathcal{K}$ is a finite union of balls contained in $\mathbb{Z}_{2}$. We
pick two non-negative, continuous functions $j(x\mid y)$, $j\left(  y\mid
x\right)  \in\mathcal{C}(\mathcal{K}\times\mathcal{K},\mathbb{R})$ (the
$\mathbb{R}$-vector space of continuous functions on $\mathcal{K}%
\times\mathcal{K}$). The function $j(y\mid x)$ gives the transition density
rate (per unit of time) from $x$ to $y$, i.e.,
\[
\mathbb{P}(x,B)=%
%TCIMACRO{\tint \limits_{B}}%
%BeginExpansion
{\textstyle\int\limits_{B}}
%EndExpansion
j(y\mid x)dy
\]
is the transition probability from $x$ into $B$ (per unit of time), where $B$
is a Borel subset of $\mathcal{K}$. In general the functions $j(x\mid y)$,
$j\left(  y\mid x\right)  $ are different. We denote by $\mathcal{C}%
(\mathcal{K},\mathbb{R})$ the $\mathbb{R}$-vector space of continuous
functions on $\mathcal{K}$, and introduce the operator $\boldsymbol{J}$
defined as%
\begin{equation}
\boldsymbol{J}f(x)=\int\limits_{\mathcal{K}}\left\{  j(x\mid y)f(y)-j(y\mid
x)f(x)\right\}  dy\text{, }f\in\mathcal{C}(\mathcal{K},\mathbb{R})\text{.}
\label{Operator_J}%
\end{equation}
By endowing $\mathcal{C}(\mathcal{K},\mathbb{R})$ with the norm $\left\Vert
f\right\Vert _{\infty}=\max_{x\in\mathcal{K}}\left\vert f\left(  x\right)
\right\vert $; space $\left(  \mathcal{C}(\mathcal{K},\mathbb{R}),\left\Vert
\cdot\right\Vert _{\infty}\right)  $ becomes a Banach space, and the mapping
$f\rightarrow\boldsymbol{J}f$ gives rise to a well-defined, linear, bounded
operator from $\mathcal{C}(\mathcal{K},\mathbb{R})$\ into itself,
\cite{Zuniga-networks}. We now assume that
\begin{equation}
j(x\mid y)\leq j(y\mid x)\text{ for any }x,y\in\mathcal{K}\text{.}
\label{Hypothesis 1}%
\end{equation}
Under the above hypothesis, the operator $\boldsymbol{J}$ satisfies the
positive maximum principle, i.e., if $h\in\mathcal{C}(\mathcal{K},\mathbb{R})$
and $\max_{x\in\mathcal{K}}h\left(  x\right)  =h(x_{0})\geq0$, then $\left(
\boldsymbol{J}h\right)  (x_{0})\leq0$. The evolution equation%
\begin{equation}
\frac{du\left(  x,\tau\right)  }{d\tau}=\int\limits_{\mathcal{K}}\left\{
j(x\mid y)u(y,\tau)-j(y\mid x)u(x,\tau)\right\}  dy\text{, }\tau\geq
0,x\in\mathcal{K}\text{,} \label{heat-equation}%
\end{equation}
is a $2$-adic heat equation, which means the following: there exists a
probability measure $p_{\tau}\left(  x,\cdot\right)  $, $t\in\left[
0,T\right]  $, with $T=T(u_{0})$, $x\in\mathcal{K}$, on the Borel $\sigma
$-algebra of $\mathcal{K}$, such that the Cauchy problem%
\[
\left\{
\begin{array}
[c]{ll}%
u\left(  \cdot,\tau\right)  \in\mathcal{C}^{1}\left(  \left[  0,T\right]
,\mathcal{C}\left(  \mathcal{K},\mathbb{R}\right)  \right)  ; & \\
& \\
\frac{du\left(  x,\tau\right)  }{d\tau}=\int\limits_{\mathcal{K}}\left\{
j(x\mid y)u(y,\tau)-j(y\mid x)u(x,\tau)\right\}  dy, & \tau\in\left[
0,T\right]  ,x\in\mathcal{K};\\
& \\
u\left(  x,0\right)  =u_{0}\left(  x\right)  \in\mathcal{C}\left(
\mathcal{K},\mathbb{R}_{+}\right)  . &
\end{array}
\right.
\]
has a unique solution of the form%
\[
u(x,\tau)=\int\limits_{\mathcal{K}}u_{0}(y)p_{\tau}\left(  x,dy\right)  .
\]
In addition, $p_{\tau}\left(  x,\cdot\right)  $ is the transition function of
a Markov process $\mathfrak{X}$ whose paths are right continuous and have no
discontinuities other than jumps; see \cite[Theorem 3.1]{Zuniga-networks}.

We now take $\mathcal{K}$ to be a disjoint union of a finite number of balls
of type $I+2^{l}\mathbb{Z}_{2}$, $I\in G_{l}^{0}$. Then, the heat equation
(\ref{heat-equation}) is the master equation of an ultrametric network, where
the states of the network are organized in $\#G_{l}^{0}$ clusters; each
cluster correspond to a ball $I+2^{l}\mathbb{Z}_{2}$. Inside of each ball the
states are organized in an infinite rooted tree. There are random transitions
between the states inside each ball and between states in different balls,
controlled by the functions $j(x\mid y)$ and $j(y\mid x)$. These networks may
have absorbing states, which means that at finite time $\tau_{0}$, the state
of the network may get trapped in \ a small neighborhood of a fixed state; in
this case, the time evolution of the network is not described by
(\ref{heat-equation}) for $\tau>\tau_{0}$; see \cite{Zuniga-networks}, for
further details.

A relevant case is obtained by taking $j(x\mid y)=j(y\mid x)=J(x-y)$, where
$J:\mathbb{Z}_{2}\rightarrow\mathbb{R}$ is a non-negative function satisfying
$\int_{\mathbb{Z}_{2}}J(x)dx=1$. From now on, we assume $J$ is integrable, but
not necessarily continuous. In this case, the $\boldsymbol{J}$ operator takes
the form
\[
\boldsymbol{J}\varphi\left(  x\right)  =J(x)\ast\varphi\left(  x\right)
-\varphi\left(  x\right)  =%
%TCIMACRO{\dint \limits_{\mathbb{Z}_{2}}}%
%BeginExpansion
{\displaystyle\int\limits_{\mathbb{Z}_{2}}}
%EndExpansion
J(x-y)\left(  \varphi\left(  y\right)  -\varphi\left(  x\right)  \right)  dy.
\]
By using that $\left\Vert J\ast\varphi\right\Vert _{\rho}\leq\left\Vert
J\right\Vert _{1}\left\Vert \varphi\right\Vert _{\rho}=\left\Vert
\varphi\right\Vert _{\rho}$, for $\varphi\in L^{\rho}\left(  \mathbb{Z}%
_{2}\right)  $, $\rho\in\left[  1,\infty\right]  $, cf. \cite[Chapter II,
Theorem 1.7]{Taibleson},%
\[
\boldsymbol{J}:L^{\rho}\left(  \mathbb{Z}_{2}\right)  \rightarrow L^{\rho
}\left(  \mathbb{Z}_{2}\right)
\]
is a well-defined linear bounded operator satisfying $\left\Vert
\boldsymbol{J}\right\Vert \leq2$.

The corresponding heat equation
\begin{equation}
\frac{\partial}{\partial\tau}u\left(  x,\tau\right)  =\boldsymbol{J}u\left(
x,\tau\right)  \text{, }x\in\mathbb{Q}_{2}\text{, }\tau\geq0, \label{Eq_5}%
\end{equation}
with initial datum $u\left(  x,0\right)  =u_{0}\left(  x\right)  \in L^{\rho
}\left(  \mathbb{Z}_{2}\right)  $ has a unique solution $u\left(
x,\tau\right)  =e^{\tau\boldsymbol{J}}u_{0}\left(  x\right)  $. The semigroup
$\left\{  e^{\tau\boldsymbol{J}}\right\}  _{\tau\geq0}$ is Feller, so there is
a Markov process attached to equation (\ref{Eq_5}), see e.g.
\cite{Zuniga-networks}, \cite{Zuniga-Anselmo}.

\subsection{$2$-Adic Schr\"{o}dinger equations coming from master equations}

We now perform a Wick rotation ($\tau=it$, $t\geq0$, with $i=\sqrt{-1}$, and
$\Psi\left(  x,t\right)  =u\left(  x,it\right)  $) in (\ref{heat-equation}) to
obtain a Schr\"{o}dinger equation. It is more convenient to change the
notation. We set $A(x,y)=j(x\mid y)$, $B\left(  x,y\right)  =j(y\mid x)$,
where $A(x,y),B\left(  x,y\right)  $ are non-negative, continuous, symmetric
functions ($A(x,y)=A\left(  y,x\right)  $, $B(x,y)=B\left(  y,x\right)  $).
With this notation, Schr\"{o}dinger equation takes the form%
\begin{align}
i\frac{\partial}{\partial t}\Psi\left(  x,t\right)   &  =-\int
\limits_{\mathcal{K}}\left\{  A(x,y)\Psi(y,t)-B\left(  x,y\right)
\Psi(x,t)\right\}  dy\label{Eq_15}\\
&  =-\int\limits_{\mathcal{K}}A(x,y)\Psi(y,t)dy+S(x)\Psi(x,t),\nonumber
\end{align}
for $t\geq0,x\in\mathcal{K}$, where%
\[
S(x):=\int\limits_{\mathcal{K}}B\left(  x,y\right)  dy\in\mathcal{C}%
(\mathcal{K},\mathbb{R}).
\]

\begin{remark}
Notice that in (\ref{Eq_15}), the mass $m$ is one. The reason is that the mass
can be absorbed by the functions $A(x,y)$, $B\left(  x,y\right)  $. The
functions $A(x,y)$, $B\left(  x,y\right)  $ are not required to satisfy the
hypothesis (\ref{Hypothesis 1}).
\end{remark}

The operators%
\begin{equation}
\Phi\left(  x\right)  \rightarrow-\int\limits_{\mathcal{K}}A(x,y)\Phi
(y)dy\text{,} \label{Operator_1}%
\end{equation}
and%
\begin{equation}
\Phi\left(  x\right)  \rightarrow S(x)\Phi(x) \label{Operator_2}%
\end{equation}
are linear bounded operators from $L^{2}(\mathcal{K})$ into itself. Indeed,
for operator (\ref{Operator_1}), by Cauchy-Schwartz inequality and the fact
that $\mu_{Haar}(\mathcal{K})=\int_{\mathcal{K}}dx<\infty$,
\[
\left\vert -\int\limits_{\mathcal{K}}A(x,y)\Phi(y)dy\right\vert \leq\left\Vert
A\right\Vert _{\infty}\int\limits_{\mathcal{K}}\left\vert \Phi(y)\right\vert
dy\leq\left\Vert A\right\Vert _{\infty}\sqrt{\mu_{Haar}(\mathcal{K}%
)}\left\Vert \Phi\right\Vert _{2},
\]
then%
\[
\left\Vert -\int\limits_{\mathcal{K}}A(x,y)\Phi(y)dy\right\Vert _{2}%
\leq\left\Vert A\right\Vert _{\infty}\text{ }\mu_{Haar}(\mathcal{K})\left\Vert
\Phi\right\Vert _{2}.
\]
For operator (\ref{Operator_2}), $\left\Vert S(x)\Phi(x)\right\Vert _{2}%
\leq\left\Vert S\right\Vert _{\infty}$ $\left\Vert \Phi\right\Vert _{2}$.

Both operators are symmetric, and then self-adjoint on $L^{2}(\mathcal{K})$.
Indeed, by Fubini's theorem,%
\begin{gather*}
\left\langle \int\limits_{\mathcal{K}}A(x,y)\Phi(y)dy,\Phi(x)\right\rangle
=\int\limits_{\mathcal{K}}\left\{  \int\limits_{\mathcal{K}}A(x,y)\Phi
(y)dy\right\}  \overline{\Phi(x)}dx\\
=\int\limits_{\mathcal{K}}\left\{  \int\limits_{\mathcal{K}}A(x,y)\overline
{\Phi(x)}dx\right\}  \Phi(y)dy=\int\limits_{\mathcal{K}}\Phi(y)\left\{
\overline{\int\limits_{\mathcal{K}}A(x,y)\Phi(x)dx}\right\}  dy\\
=\left\langle \Phi(x),\int\limits_{\mathcal{K}}A(x,y)\Phi(y)dy\right\rangle .
\end{gather*}
The case of operator (\ref{Operator_2}) is treated in a \ similar form.

So, the operator%
\[
\Psi\left(  x,t\right)  \rightarrow-\int\limits_{\mathcal{K}}\left\{
A(x,y)\Psi(y,t)-B\left(  x,y\right)  \Psi(x,t)\right\}  dy,
\]
for $t\geq0$, is self-adjoint on $L^{2}(\mathcal{K})$.

It is convenient to rewrite equation (\ref{Eq_15}) as%
\begin{align}
i\frac{\partial}{\partial t}\Psi\left(  x,t\right)   &  =-\int
\limits_{\mathcal{K}}A(x,y)\left\{  \Psi(y,t)-\Psi(x,t)\right\}
dy+V(x)\Psi\left(  x,t\right) \label{Eq_SChrodinger_4}\\
&  =:\boldsymbol{H}\Psi\left(  x,t\right)  =:\left(  -\boldsymbol{H}%
_{0}+\boldsymbol{V}\right)  \Psi\left(  x,t\right)  ,\nonumber
\end{align}
where%
\begin{equation}
V(x):=\int\limits_{\mathcal{K}}\left\{  B(x,y)-A\left(  x,y\right)  \right\}
dy\in\mathcal{C}\left(  \mathcal{K}_{l},\mathbb{R}\right)  , \label{Potential}%
\end{equation}
and $\boldsymbol{V}$ is operator of multiplication by $V(x)$.

Now since $\boldsymbol{H}$ is self-adjoint on $L^{2}(\mathcal{K})$, by Stone's
theorem on one-parameter unitary groups, there exists a one-paremeter family
of unitary operators $\left\{  e^{-it\boldsymbol{H}}\right\}  _{t\geq0}$, such
that $\Psi\left(  x,t\right)  =e^{-it\boldsymbol{H}}\Psi_{0}\left(  x\right)
$ is the unique solution of the Cauchy problem%
\begin{equation}
\left\{
\begin{array}
[c]{l}%
\Psi\left(  \cdot,t\right)  \in L^{2}(\mathcal{K})\text{, }t\geq0;\text{ }%
\Psi\left(  x,\cdot\right)  \in\mathcal{C}^{1}(\mathbb{R}_{+})\text{, }%
x\in\mathcal{K}\\
\\
i\frac{\partial}{\partial t}\Psi\left(  x,t\right)  =\boldsymbol{H}\Psi\left(
x,t\right)  \text{, }x\in\mathcal{K},t\geq0\\
\\
\Psi\left(  x,0\right)  =\Psi_{0}\left(  x\right)  \in L^{2}(\mathcal{K}).
\end{array}
\right.  \label{Eq_Cauchy_problem-2}%
\end{equation}

\begin{remark}
Let $V:\mathbb{Z}_{2}\rightarrow\mathbb{R}$ be a time-independent continuous
potential, and let $\boldsymbol{V}$ be the operator of multiplication by
$V(x)$. The $2$-adic Schr\"{o}dinger equation with time-independent potential
$V$ corresponding to (\ref{Eq_5}) is%
\begin{equation}
i\frac{\partial\Psi(x,t)}{\partial t}=\left(  -m\boldsymbol{J}+\boldsymbol{V}%
\right)  \Psi(x,t),\text{ }x\in\mathbb{Z}_{2},t\geq0\text{,} \label{Eq_5A}%
\end{equation}
where $m>0$ is the mass of the particle. The operator $-m\boldsymbol{J}%
+\boldsymbol{V}$ is self-adjoint on $L^{2}(\mathbb{Z}_{2})$, and by Stone's
theorem on one-parameter unitary groups, there exists a one-paremeter family
of unitary operators $\left\{  e^{-it\left(  -m\boldsymbol{J}+\boldsymbol{V}%
\right)  }\right\}  _{t\geq0}$ on $L^{2}(\mathbb{Z}_{2})$, such that
$\Psi\left(  x,t\right)  =e^{-it\left(  -m\boldsymbol{J}+\boldsymbol{V}%
\right)  }\Psi_{0}\left(  x\right)  $, $\Psi_{0}\left(  x\right)  =\Psi\left(
x,0\right)  $, is the solution of the Cauchy problem associated with
(\ref{Eq_5A}).
\end{remark}

\subsection{\label{Sub_CTQMC and Born Rule}Construction\ CTQMCs via Born rule}

We now take $\mathcal{K}=%
%TCIMACRO{\tbigsqcup \nolimits_{I\in G_{l}^{0}}}%
%BeginExpansion
{\textstyle\bigsqcup\nolimits_{I\in G_{l}^{0}}}
%EndExpansion
\left(  I+2^{l}\mathbb{Z}_{2}\right)  $, where $G_{l}^{0}$ is a finite subset
of $\mathbb{Z}_{2}$, $\Psi_{I}\left(  x\right)  :=2^{\frac{l}{2}}\Omega\left(
2^{l}\left\vert x-I\right\vert _{2}\right)  $, with $\Omega\left(
2^{l}\left\vert x-I\right\vert _{2}\right)  $ denoting the characteristic
function of the ball $I+2^{l}\mathbb{Z}_{2}$, and $\Psi\left(  x,t\right)
=e^{-it\boldsymbol{H}}\Psi_{I}\left(  x\right)  $ as before. Notice that
\[
1=\left\Vert \Psi_{I}\left(  x\right)  \right\Vert _{2}=\left\Vert \Psi\left(
x,t\right)  \right\Vert _{2}=\sqrt{%
%TCIMACRO{\dint \limits_{\mathcal{K}}}%
%BeginExpansion
{\displaystyle\int\limits_{\mathcal{K}}}
%EndExpansion
\left\vert \Psi\left(  x,t\right)  \right\vert ^{2}dx}\text{;}%
\]
then, by Born's rule,
\[%
%TCIMACRO{\dint \limits_{B}}%
%BeginExpansion
{\displaystyle\int\limits_{B}}
%EndExpansion
\left\vert \Psi\left(  x,t\right)  \right\vert ^{2}dx
\]
gives the probability of finding the system in a state supported in
$B\subset\mathcal{K}$ (a Borel subset) given that at time zero the state of
the system was given by $\Psi_{I}\left(  x\right)  $. Therefore,%
\begin{equation}
\widetilde{\pi}_{_{J,I}}\left(  t\right)  =%
%TCIMACRO{\dint \limits_{J+2^{l}\mathbb{Z}_{2}}}%
%BeginExpansion
{\displaystyle\int\limits_{J+2^{l}\mathbb{Z}_{2}}}
%EndExpansion
\left\vert \Psi\left(  x,t\right)  \right\vert ^{2}dx \label{Markov-1}%
\end{equation}
is a transition probability between a state supported in the ball
$I+2^{l}\mathbb{Z}_{2}$ to a state supported in the ball $J+2^{l}%
\mathbb{Z}_{2}$ at the time $t$. Notice that%
\begin{equation}%
%TCIMACRO{\dsum \limits_{J\in G_{l}^{0}}}%
%BeginExpansion
{\displaystyle\sum\limits_{J\in G_{l}^{0}}}
%EndExpansion
\widetilde{\pi}_{_{J,I}}\left(  t\right)  =1. \label{Markov-2}%
\end{equation}
Then, if we identify the ball $I+2^{l}\mathbb{Z}_{2}$ with vertex $I\in
G_{l}^{0}$ of a complete graph, the matrix $\left[  \widetilde{\pi}_{_{J,I}%
}\left(  t\right)  \right]  $ defines a quantum Markov chain ( CTQMC). This
approach was introduced in \cite{Zuniga-Mayes}. The drawback of this approach
is that it requires the solution of Cauchy problem (\ref{Eq_Cauchy_problem-2}).

\section{\label{Sect_4}Discretizations of $2$-adic Schr\"{o}dinger equations
and quantum networks}

By the discretization of a function, we mean an approximation of it, which is
a function of a finite-dimensional vector space. By the discretization of an
operator, we mean an approximation of it by a matrix acting on a
finite-dimensional vector space. Well-known techniques exist for approximating
abstract evolution equations. Our approach is highly influenced by
\cite[Section 5.4]{Milan}. The goal of this section is to construct space
approximations of (\ref{Eq_Cauchy_problem-2}). Such an approximation is a
system of ordinary differential equations in
\[
\left[  \Psi_{I}^{\left(  l\right)  }\left(  t\right)  \right]  _{I\in G_{l}%
}\in\mathbb{C}^{\#G_{l}},
\]
where $G_{l}$ is a finite subset, and $l$ is positive integer. The finite set
$G_{l}$ is a space discretization of a completely disconnected space
$(\mathbb{Z}_{2})$ such that in the limit when $l\rightarrow\infty$, $G_{l}$
becomes $\mathbb{Z}_{2}$.

\subsection{The space $\mathcal{D}_{l}\left(  \mathbb{Z}_{2}\right)  $}

We now construct approximations for the solution $\Psi\left(  x,t\right)  $ of
(\ref{Eq_Cauchy_problem-2}). The $\mathbb{C}$-vector space $\mathcal{D}%
_{l}\left(  \mathbb{Z}_{2}\right)  $, $l\in\mathbb{N}$, consists of the test
functions supported in the ball $\mathbb{Z}_{2}$\ of the form%
\[
\varphi^{\left(  l\right)  }\left(  x\right)  =%
%TCIMACRO{\dsum \limits_{I\in G_{l}}}%
%BeginExpansion
{\displaystyle\sum\limits_{I\in G_{l}}}
%EndExpansion
\varphi_{I}2^{\frac{l}{2}}\Omega\left(  2^{l}\left\vert x-I\right\vert
_{2}\right)  \text{, }\varphi_{I}\in\mathbb{C}\text{,}%
\]
where $\Omega\left(  2^{l}\left\vert x-I\right\vert _{2}\right)  $ is the
characteristic function of the ball $I+2^{l}\mathbb{Z}_{2}$, $G_{l}%
=\mathbb{Z}_{2}/2^{l}\mathbb{Z}_{2}\simeq\mathbb{Z}/2^{l}\mathbb{Z}$ is the
set of binary sequences of $l$-bits, i.e., $I=\sum_{k=0}^{l-1}I_{k}2^{k}$ ,
with $I_{k}\in\left\{  0,1\right\}  $. Notice that the cardinality of $G_{l}$
is $2^{l}$.

Given $\varphi^{\left(  l\right)  }\left(  x\right)  ,\theta^{\left(
l\right)  }\left(  x\right)  \in\mathcal{D}_{l}\left(  \mathbb{Z}_{2}\right)
$, we recall that%
\[
\left\langle \varphi^{\left(  l\right)  }\left(  x\right)  ,\theta^{\left(
l\right)  }\left(  x\right)  \right\rangle =%
%TCIMACRO{\dint \limits_{\mathbb{Z}_{2}}}%
%BeginExpansion
{\displaystyle\int\limits_{\mathbb{Z}_{2}}}
%EndExpansion
\varphi^{\left(  l\right)  }\left(  x\right)  \overline{\theta^{\left(
l\right)  }\left(  x\right)  }dx\text{,}%
\]
is restriction of the standard inner product of $L^{2}(\mathbb{Z}_{2})$ to
$\mathcal{D}_{l}\left(  \mathbb{Z}_{2}\right)  $. The set of vectors%
\begin{equation}
\left\{  2^{\frac{l}{2}}\Omega\left(  2^{l}\left\vert x-I\right\vert
_{2}\right)  \right\}  _{I\in G_{l}} \label{Eq_basis}%
\end{equation}
is an orthonormal basis for $\mathcal{D}_{l}\left(  \mathbb{Z}_{2}\right)  $.
By identifying $\varphi^{\left(  l\right)  }\left(  x\right)  $ with the
column vector $\left[  \varphi_{I}\right]  _{I\in G_{l}}\in\mathbb{C}^{2^{l}}%
$, we obtain an isometry between $\mathcal{D}_{l}\left(  \mathbb{Z}%
_{2}\right)  $ and the Hilbert space $\mathbb{C}^{2^{l}}$; this last space is
the standard Hilbert space used in quantum computing.

We recall that $\mathcal{D}\left(  \mathbb{Z}_{2}\right)  =\cup_{l\in
\mathbb{N}}\mathcal{D}_{l}\left(  \mathbb{Z}_{2}\right)  $ and that
$\mathcal{D}\left(  \mathbb{Z}_{2}\right)  \subset\mathcal{C}(\mathbb{Z}_{2}%
)$, where $\mathcal{C}(\mathbb{Z}_{2})$ is the $\mathbb{C}$-vector space of
continuous functions defined on $\mathbb{Z}_{2}$. Furthermore, $\mathcal{D}%
\left(  \mathbb{Z}_{2}\right)  $ is dense in $L^{1}(\mathbb{Z}_{2})$, see e.g.
\cite[Section 4.3]{Alberio et al}. Now, the fact that the Haar measure of
$\mathbb{Z}_{2}$ is one, and the Cauchy--Schwarz inequality imply that%
\[
L^{1}(\mathbb{Z}_{2})\supset L^{2}(\mathbb{Z}_{2})\supset L^{\infty
}(\mathbb{Z}_{2})\supset\mathcal{C}(\mathbb{Z}_{2}),\supset\mathcal{D}\left(
\mathbb{Z}_{2}\right)  ,
\]
where all the embeddings are continuous, more precisely,%
\[
\left\Vert f\right\Vert _{\infty}\geq\left\Vert f\right\Vert _{2}%
\geq\left\Vert f\right\Vert _{1}.
\]
Therefore, the space $\mathcal{D}_{l}\left(  \mathbb{Z}_{2}\right)  $ is dense
in $L^{\rho}(\mathbb{Z}_{2})$ for $\rho\in\left\{  1,2,\infty\right\}  $:
given $f\in L^{\rho}(\mathbb{Z}_{2})$ and $\epsilon>0$, there exists
$l=l\left(  \epsilon\right)  $ and $f^{\left(  l\right)  }\in\mathcal{D}%
_{l}\left(  \mathbb{Z}_{2}\right)  $ of the form
\begin{equation}
f^{\left(  l\right)  }(x)=%
%TCIMACRO{\dsum \limits_{I\in G_{l}}}%
%BeginExpansion
{\displaystyle\sum\limits_{I\in G_{l}}}
%EndExpansion
f_{I}^{\left(  l\right)  }2^{\frac{l}{2}}\Omega\left(  2^{l}\left\vert
x-I\right\vert _{2}\right)  \label{Approx_f_l}%
\end{equation}
such that $\left\Vert f-f^{\left(  l\right)  }\right\Vert _{\rho}<\epsilon$.
This result does not give an algorithm for estimating the coefficients
$f_{I}^{\left(  l\right)  }$.

Let $f:\mathbb{Z}_{2}\rightarrow\mathbb{C}$ be an integrable function. We
extend $f$ outside of the unit ball as zero. The average of the function $f$
in ball $B_{-l}(x_{0})=\left\{  z\in\mathbb{Z}_{2};\left\vert z-x_{0}%
\right\vert _{2}\leq2^{-l}\right\}  $ is given by
\[
\mathrm{Aver}(f,B_{-l}(x_{0})):=2^{l}%
%TCIMACRO{\dint \limits_{\left\vert z-x_{0}\right\vert _{2}\leq2^{-l}}}%
%BeginExpansion
{\displaystyle\int\limits_{\left\vert z-x_{0}\right\vert _{2}\leq2^{-l}}}
%EndExpansion
f(z)dz,
\]
where the volume (the Haar measure) of $B_{-l}(x_{0})$\ is $2^{-l}$. The
integrability condition implies the existence of subset $\mathcal{N}%
\subset\mathbb{Z}_{2}$ with Haar measure zero, such that
\[
\lim_{l\rightarrow\infty}\mathrm{Aver}(f,B_{-l}(x_{0}))=f(x_{0})\text{ for any
}x_{0}\in\mathbb{Z}_{2}\smallsetminus\mathcal{N}\text{, }%
\]
see \cite[Chap. II, Theorem 1.14]{Taibleson}. We take $f^{\left(  l\right)
}(x)$ as (\ref{Approx_f_l}), with $f_{I}^{\left(  l\right)  }2^{\frac{l}{2}%
}=\mathrm{Aver}(f,B_{-l}(I))$. Then, assuming that $x\in B_{-l}(I)$,%
\[
\left\Vert f\left(  x\right)  -f^{\left(  l\right)  }(x)\right\Vert _{1}%
\leq\left\Vert f\left(  x\right)  -f^{\left(  l\right)  }(x)\right\Vert
_{\infty}=\left\Vert f\left(  x\right)  -\mathrm{Aver}(f,B_{-l}(I))\right\Vert
_{\infty}\rightarrow0\text{ as }l\rightarrow\infty,
\]
for any $x\in\mathbb{Z}_{2}$ outside of a measure zero set. In the
verification of this last approximation, we may assume that $f$ is continuous,
since the continuous function are dense in $L^{1}(\mathbb{Z}_{2})$.

\subsubsection{Approximations of wavefunctions}

We now construct approximations $\Psi^{\left(  l\right)  }\left(  x,t\right)
$ to the solutions of (\ref{Eq_SChrodinger_4}) and (\ref{Eq_5A}),
\cite[Section 5.4]{Milan}. Since $\Psi\left(  \cdot,t\right)  \in
L^{2}(\mathbb{Z}_{2})$, by using the density of $\mathcal{D}_{l}\left(
\mathbb{Z}_{2}\right)  $ in $L^{2}(\mathbb{Z}_{2})\subset L^{1}(\mathbb{Z}%
_{2})$, we have an approximation of the form%
\[
\Psi^{\left(  l\right)  }\left(  x,t\right)  =%
%TCIMACRO{\dsum \limits_{I\in G_{l}}}%
%BeginExpansion
{\displaystyle\sum\limits_{I\in G_{l}}}
%EndExpansion
\Psi_{I}^{\left(  l\right)  }\left(  t\right)  2^{\frac{l}{2}}\Omega\left(
2^{l}\left\vert x-I\right\vert _{2}\right)  ,
\]
where $\Psi_{I}^{\left(  l\right)  }\left(  t\right)  $ is differentiable,
i.e., $\Psi_{I}^{\left(  l\right)  }\left(  t\right)  \in\mathcal{C}%
^{1}\left(  \mathbb{R}\right)  $.

Define the linear operator%
\begin{equation}%
\begin{array}
[c]{cccc}%
\boldsymbol{P}_{l}: & L^{1}(\mathbb{Z}_{2}) & \rightarrow & \mathcal{D}%
_{l}\left(  \mathbb{Z}_{2}\right) \\
&  &  & \\
& \phi\left(  x\right)  & \rightarrow & \phi^{\left(  l\right)  }\left(
x\right)  =%
%TCIMACRO{\dsum \limits_{I\in G_{l}}}%
%BeginExpansion
{\displaystyle\sum\limits_{I\in G_{l}}}
%EndExpansion
\phi_{I}^{\left(  l\right)  }2^{\frac{l}{2}}\Omega\left(  2^{l}\left\vert
x-I\right\vert _{2}\right)  ,
\end{array}
\label{Eq_P_l}%
\end{equation}
where $\phi_{I}^{\left(  l\right)  }2^{\frac{l}{2}}=\mathrm{Aver}%
(f,B_{-l}(I))$. It is not difficult to verify that $\left\Vert \boldsymbol{P}%
_{l}\phi\right\Vert _{2}\leq\left\Vert \phi\right\Vert _{2}$, i.e., operator
$\boldsymbol{P}_{l}$\ is continuous.

The approximations $\boldsymbol{H}_{0}^{\left(  l\right)  },\boldsymbol{V}%
^{\left(  l\right)  }$ of the operators $\boldsymbol{H}_{0},\boldsymbol{V}$
are defined as follows:%
\begin{equation}
\boldsymbol{H}_{0}^{\left(  l\right)  }:\mathcal{D}_{l}\left(  \mathbb{Z}%
_{2}\right)  \text{ \ }\underrightarrow{\boldsymbol{H}_{0}}\text{ \ }%
L^{2}(\mathbb{Z}_{2})\text{ \ }\underrightarrow{\boldsymbol{P}_{l}}\text{
\ }\mathcal{D}_{l}\left(  \mathbb{Z}_{2}\right)  , \label{Distretization_1}%
\end{equation}%
\begin{equation}
\boldsymbol{V}^{\left(  l\right)  }:\mathcal{D}_{l}\left(  \mathbb{Z}%
_{2}\right)  \text{ \ }\underrightarrow{\boldsymbol{V}}\text{ \ }%
L^{2}(\mathbb{Z}_{2})\text{ \ }\underrightarrow{\boldsymbol{P}_{l}}\text{
\ }\mathcal{D}_{l}\left(  \mathbb{Z}_{2}\right)  . \label{Distretization_2}%
\end{equation}
Notice that if $V\in L^{1}(\mathbb{Z}_{2})$ and $\varphi\in L^{2}%
(\mathbb{Z}_{2})$, then $V\varphi\in L^{2}(\mathbb{Z}_{2})$, by the
Cauchy-Schwartz inequality.

Now, $\boldsymbol{H}_{0}^{\left(  l\right)  },\boldsymbol{V}^{\left(
l\right)  }$ are linear operators on a finite dimensional Hilbert space
$\mathcal{D}_{l}\left(  \mathbb{Z}_{2}\right)  $; by using basis
(\ref{Eq_basis}), these operators are represented by matrices of size
$2^{l}\times2^{l}$ acting of column vectors%
\[
\left[  \Psi_{I}^{\left(  l\right)  }\left(  t\right)  \right]  :=\left[
\Psi_{I}^{\left(  l\right)  }\left(  t\right)  \right]  _{I\in G_{l}}.
\]
Set $\boldsymbol{H}^{\left(  l\right)  }=-\boldsymbol{H}_{0}^{\left(
l\right)  }+\boldsymbol{V}^{\left(  l\right)  }$, for (\ref{Eq_SChrodinger_4}%
), and $\boldsymbol{H}_{0}^{\left(  l\right)  }=-m\boldsymbol{J}^{\left(
l\right)  }+\boldsymbol{V}^{\left(  l\right)  }$, for (\ref{Eq_5A}),. Then,
the discretization of the systems (\ref{Eq_SChrodinger_4}) and (\ref{Eq_5A}%
)\ is
\begin{equation}
\left\{
\begin{array}
[c]{l}%
\left[  \Psi_{I}^{\left(  l\right)  }\left(  t\right)  \right]  \in
\mathbb{C}^{2^{l}}\text{, }t\geq0;\text{ }\Psi_{I}^{\left(  l\right)  }\left(
t\right)  \in C^{1}(\mathbb{R}_{+})\text{, }I\in G_{l}\\
\\
i\frac{\partial}{\partial t}\left[  \Psi_{I}^{\left(  l\right)  }\left(
t\right)  \right]  =\boldsymbol{H}^{\left(  l\right)  }\left[  \Psi
_{I}^{\left(  l\right)  }\left(  t\right)  \right]  \text{, }I\in G_{l}%
,t\geq0\\
\\
\left[  \Psi_{I}^{\left(  l\right)  }\left(  0\right)  \right]  =\left[
\Psi_{I}^{\left(  \text{init}\right)  }\right]  \in\mathbb{C}^{2^{l}}.
\end{array}
\right.  \label{Eq_4}%
\end{equation}

\begin{remark}
$\boldsymbol{V}^{\left(  l\right)  }$ is the matrix of the linear operator
$\varphi^{\left(  l\right)  }\left(  x\right)  \rightarrow\varphi^{\left(
l\right)  }\left(  x\right)  V^{\left(  l\right)  }\left(  x\right)  $,
$\varphi^{\left(  l\right)  }\in\mathcal{D}_{l}\left(  \mathbb{Z}_{2}\right)
$, where $V^{\left(  l\right)  }$ is the approximation of $V$\ given by
\[
V^{\left(  l\right)  }\left(  x\right)  =%
%TCIMACRO{\dsum \limits_{I\in G_{l}}}%
%BeginExpansion
{\displaystyle\sum\limits_{I\in G_{l}}}
%EndExpansion
V_{I}^{\left(  l\right)  }2^{\frac{l}{2}}\Omega\left(  2^{l}\left\vert
x-I\right\vert _{2}\right)  ,
\]
with $V_{I}^{\left(  l\right)  }2^{\frac{l}{2}}=\mathrm{Aver}(V,B_{-l}(I))$.
Therefore, $\boldsymbol{V}^{\left(  l\right)  }=\mathrm{diag}\left[
2^{\frac{-l}{2}}\mathrm{Aver}(V,B_{-l}(I))\right]  _{I\in G_{l}}$.
\end{remark}

\subsection{\label{Sub_Quantum_networks}Quantum networks}

Our next step is to show that the discrete $2$-adic Schr\"{o}dinger equations
(\ref{Eq_SChrodinger_4}) and (\ref{Eq_5A}) describe CTQMCs, and that some
equations of type (\ref{Eq_5A}) describe the Farhi-Gutmann CTQWs,
\cite{Farhi-Gutman}, see also \cite{Mulkne-Blumen}. Here, we review the
construction of CTQWs\ following \cite{Mulkne-Blumen}, \cite{Venegas-Andraca}.

From now on, $\mathcal{H}$ denotes the Hilbert space $\mathbb{C}^{2^{l}}$,
with norm $\left\Vert \cdot\right\Vert $, and canonical basis as $\left\{
\left\vert e_{I}\right\rangle \right\}  _{I\in G_{l}}$. Here, use use the
Dirac notation. We assume that $\boldsymbol{H}^{\left(  l\right)  }$ is a
Hermitian matrix so $\exp(-it\boldsymbol{H}^{\left(  l\right)  })$ is unitary
matrix. We identify $G_{l}$ with an graph with vertices $I\in G_{l}$. We
define the transition probability $\pi_{I,J}\left(  t\right)  $ from $J$ to
$I$ as
\[
\pi_{I,J}\left(  t\right)  =\left\vert \left\langle e_{I}\right\vert
e^{-it\boldsymbol{H}^{\left(  l\right)  }}\left\vert e_{J}\right\rangle
\right\vert ^{2}\text{, for }J,I\in G_{l}.
\]
If $\pi_{I,J}\left(  t\right)  \neq\pi_{_{J,I}}\left(  t\right)  $ there is an
oriented edge from $J$ to $I$. If $\pi_{I,J}\left(  t\right)  =\pi_{_{J,I}%
}\left(  t\right)  $, there is only an unoriented edge between $J$ and $I$.

Note that%
\[%
%TCIMACRO{\dsum \limits_{I\in G_{l}}}%
%BeginExpansion
{\displaystyle\sum\limits_{I\in G_{l}}}
%EndExpansion
\pi_{I,J}\left(  t\right)  =%
%TCIMACRO{\dsum \limits_{I\in G_{l}}}%
%BeginExpansion
{\displaystyle\sum\limits_{I\in G_{l}}}
%EndExpansion
\left\vert \left\langle e_{I}\right\vert e^{-it\boldsymbol{H}^{\left(
l\right)  }}\left\vert e_{J}\right\rangle \right\vert ^{2}=1.
\]
Indeed, using that%
\[
\Theta_{J}\left(  t\right)  =:e^{-it\boldsymbol{H}^{\left(  l\right)  }%
}\left\vert e_{J}\right\rangle =%
%TCIMACRO{\dsum \limits_{K\in G_{l}}}%
%BeginExpansion
{\displaystyle\sum\limits_{K\in G_{l}}}
%EndExpansion
C_{K,J}\left(  t\right)  \left\vert e_{K}\right\rangle
\]
and
\[
\left\Vert \Theta_{J}\left(  t\right)  \right\Vert =\left\Vert
%TCIMACRO{\dsum \limits_{K\in G_{l}}}%
%BeginExpansion
{\displaystyle\sum\limits_{K\in G_{l}}}
%EndExpansion
C_{K,J}\left(  t\right)  \left\vert e_{K}\right\rangle \right\Vert =\sqrt{%
%TCIMACRO{\dsum \limits_{K\in G_{l}}}%
%BeginExpansion
{\displaystyle\sum\limits_{K\in G_{l}}}
%EndExpansion
\left\vert C_{K,J}\left(  t\right)  \right\vert ^{2}}=\left\Vert \left\vert
e_{J}\right\rangle \right\Vert =1
\]
since $e^{-it\boldsymbol{H}^{\left(  l\right)  }}$ is a unitary matrix, we
have%
\[%
%TCIMACRO{\dsum \limits_{I\in G_{l}}}%
%BeginExpansion
{\displaystyle\sum\limits_{I\in G_{l}}}
%EndExpansion
\left\vert \left\langle e_{I}\right\vert e^{-it\boldsymbol{H}^{\left(
l\right)  }}\left\vert e_{J}\right\rangle \right\vert ^{2}=%
%TCIMACRO{\dsum \limits_{I\in G_{l}}}%
%BeginExpansion
{\displaystyle\sum\limits_{I\in G_{l}}}
%EndExpansion
\left\vert C_{I,J}\left(  t\right)  \right\vert ^{2}=1.
\]
We call to the continuous-time Markov chain on $G_{l}$ determined by the
transition probabilities $\left[  \pi_{I,J}\left(  t\right)  \right]  _{I,J\in
G_{l}}$, the quantum network associated with the discrete $2$-adic
Schr\"{o}dinger equation (\ref{Eq_4}). This construction works if we replace
$G_{l}$ with a subset $G_{l}^{0}$ of it.

The discrete Hamiltonians $\boldsymbol{H}^{\left(  l\right)  }$ depend on many
parameters, for instance, the functions $A(x,y)$, $B(x,y)$, $J(x-y)$, and the
potential $V(x)$, so the above basic construction gives rise infinitely many
different CTQMCs. The next step is to compute the matrices $\boldsymbol{H}%
^{\left(  l\right)  }$ and verify their hermiticity for a large class of parameters.

\section{\label{Sect_5}CTQWs on graphs}

The CTQWs on undirected graphs play a central role in quantum computing. In
this section, we show that this type of CTQWs can be obtained from a suitable
$2$-adic Schr\"{o}dinger equation of type (\ref{Eq_5A}). Let $\mathcal{G}$ be
an undirected, finite graph with vertices $I\in G_{l}^{0}\subset G_{l}$, and
adjacency matrix $\left[  A_{JI}\right]  _{J,I\in G_{N}^{0}}$, with
\[
A_{JI}:=\left\{
\begin{array}
[c]{ll}%
1 & \text{if the vertices }J\text{ and }I\text{ are connected}\\
& \\
0 & \text{otherwise.}%
\end{array}
\right.
\]
This matrix is symmetric, and the diagonal entries are not necessary zero,
which means that graph $\mathcal{G}$ may have self-loops.

From now on, we fix and $l$ such that $\#G_{l}^{0}\leq2^{l}$, and set
\begin{equation}
\mathcal{K}=\mathcal{K}_{l}:=%
%TCIMACRO{\tbigsqcup \limits_{I\in G_{l}^{0}}}%
%BeginExpansion
{\textstyle\bigsqcup\limits_{I\in G_{l}^{0}}}
%EndExpansion
\left(  I+2^{l}\mathbb{Z}_{2}\right)  \text{,} \label{k-l-Set}%
\end{equation}
which is an open compact subset of $\mathbb{Z}_{2}$. We also define%
\begin{equation}
J^{\left(  l\right)  }(x,y)=2^{l}%
%TCIMACRO{\tsum \limits_{J\in G_{l}^{0}}}%
%BeginExpansion
{\textstyle\sum\limits_{J\in G_{l}^{0}}}
%EndExpansion%
%TCIMACRO{\tsum \limits_{K\in G_{l}^{0}}}%
%BeginExpansion
{\textstyle\sum\limits_{K\in G_{l}^{0}}}
%EndExpansion
A_{JK}\Omega\left(  2^{l}\left\vert x-J\right\vert _{p}\right)  \Omega\left(
2^{l}\left\vert y-K\right\vert _{p}\right)  \text{,} \label{Eq_J_l}%
\end{equation}
$x$, $y\in\mathbb{Z}_{2}$, where $\left[  A_{JI}\right]  _{J,I\in G_{l}^{0}}$
is the adjacency matrix of graph $\mathcal{G}$. Notice that $J^{\left(
l\right)  }(x,y)$ is a real-valued test function on $\mathcal{K}_{l}%
\times\mathcal{K}_{l}$. We now introduce the linear operator%
\[
\boldsymbol{J}_{\mathcal{G}}\varphi\left(  x\right)  :=%
%TCIMACRO{\tint \limits_{\mathcal{K}_{l}}}%
%BeginExpansion
{\textstyle\int\limits_{\mathcal{K}_{l}}}
%EndExpansion
\left\{  \varphi\left(  y\right)  -\varphi\left(  x\right)  \right\}
J^{\left(  l\right)  }(x,y)dy\text{, for }\varphi\in C\left(  \mathcal{K}%
_{l}\right)  \text{.}%
\]
This operator extends to linear bounded operator in $L^{2}(\mathcal{K}_{l})$.

We denote by $\mathcal{X}_{l}\left(  \mathbb{Z}_{2}\right)  \subset
\mathcal{D}_{l}\left(  \mathbb{Z}_{2}\right)  $, the $\mathbb{C}$-vector space
consisting of all the test functions supported in $\mathcal{K}_{l}$\ having
the form%
\begin{equation}
\varphi\left(  x\right)  =%
%TCIMACRO{\tsum \limits_{J\in G_{l}^{0}}}%
%BeginExpansion
{\textstyle\sum\limits_{J\in G_{l}^{0}}}
%EndExpansion
\varphi_{J}2^{\frac{l}{2}}\Omega\left(  2^{l}\left\vert x-J\right\vert
_{2}\right)  \text{,} \label{Eq_X_l}%
\end{equation}
where $\varphi_{J}\in\mathbb{C}$. Notice that $\boldsymbol{J}_{\mathcal{G}%
}:\mathcal{X}_{l}\left(  \mathbb{Z}_{2}\right)  \rightarrow\mathcal{X}_{l}$
$\left(  \mathbb{Z}_{2}\right)  $ is a linear bounded operator satisfying
$\left\Vert \boldsymbol{J}_{\mathcal{G}}\right\Vert \leq2\gamma_{_{\mathcal{G}%
}}$, where $\gamma_{_{\mathcal{G}}}:=\max_{I\in G_{l}^{0}}\gamma_{I}$, with
$\gamma_{I}:=\sum_{J\in G_{l}^{0}}A_{IJ}$. We set%
\[
\mathrm{val}(I):=\left\{
\begin{array}
[c]{lll}%
\gamma_{I} & \text{if} & A_{II}=0\\
&  & \\
\gamma_{I}-1 & \text{if} & A_{II}=1.
\end{array}
\right.
\]
We call $\mathrm{val}(I)$, the valence of $I$, it is the number of connections
from $I$ to its other vertices.

By seek of simplicity, we assume that%
\[
V\left(  x\right)  =%
%TCIMACRO{\tsum \limits_{J\in G_{l}^{0}}}%
%BeginExpansion
{\textstyle\sum\limits_{J\in G_{l}^{0}}}
%EndExpansion
V_{J}2^{\frac{l}{2}}\Omega\left(  2^{l}\left\vert x-J\right\vert _{2}\right)
\text{, }V_{J}\in\mathbb{R}\text{.}%
\]
The discretization of (\ref{Eq_5A}) is obtained by computing the matrix of
$\left.  \boldsymbol{J}_{\mathcal{G}}\right\vert _{\mathcal{X}_{l}}$ assuming
that
\[
\Psi^{\left(  l\right)  }\left(  x,t\right)  =%
%TCIMACRO{\dsum \limits_{I\in G_{l}^{0}}}%
%BeginExpansion
{\displaystyle\sum\limits_{I\in G_{l}^{0}}}
%EndExpansion
\Psi_{I}^{\left(  l\right)  }\left(  t\right)  2^{\frac{l}{2}}\Omega\left(
2^{l}\left\vert x-I\right\vert _{2}\right)  .
\]
We identify $\Psi^{\left(  l\right)  }\left(  x,t\right)  $\ with the column
vector $\left[  \Psi_{I}^{\left(  l\right)  }\left(  t\right)  \right]  $. In
the computation of the matrix of $\left.  \boldsymbol{J}_{\mathcal{G}%
}\right\vert _{\mathcal{X}_{l}}$, we use that%
\[
\Omega\left(  2^{l}\left\vert z-I\right\vert _{2}\right)  \Omega\left(
2^{l}\left\vert z-J\right\vert _{2}\right)  =\left\{
\begin{array}
[c]{lll}%
0 & \text{if} & I\neq J\\
&  & \\
\Omega\left(  2^{l}\left\vert z-J\right\vert _{2}\right)  & \text{if} & I=J,
\end{array}
\right.
\]
and $2^{\frac{l}{2}}\int_{\mathbb{Z}_{2}}\Omega\left(  p^{l}\left\vert
y-K\right\vert _{p}\right)  dy=1$. The calculation is as follows:%
\begin{gather*}
\boldsymbol{J}_{\mathcal{G}}\left(  2^{\frac{l}{2}}\Omega\left(
2^{l}\left\vert x-I\right\vert _{2}\right)  \right)  =%
%TCIMACRO{\tsum \limits_{J\in G_{l}^{0}}}%
%BeginExpansion
{\textstyle\sum\limits_{J\in G_{l}^{0}}}
%EndExpansion
A_{JI}2^{\frac{l}{2}}\Omega\left(  2^{l}\left\vert x-J\right\vert _{2}\right)
-\left(
%TCIMACRO{\tsum \limits_{K\in G_{l}^{0}}}%
%BeginExpansion
{\textstyle\sum\limits_{K\in G_{l}^{0}}}
%EndExpansion
A_{IK}\right)  2^{\frac{l}{2}}\Omega\left(  2^{l}\left\vert x-I\right\vert
_{p}\right) \\
=%
%TCIMACRO{\tsum \limits_{J\in G_{l}^{0}}}%
%BeginExpansion
{\textstyle\sum\limits_{J\in G_{l}^{0}}}
%EndExpansion
A_{JI}2^{\frac{l}{2}}\Omega\left(  2^{l}\left\vert x-J\right\vert _{2}\right)
-\gamma_{I}2^{\frac{l}{2}}\Omega\left(  2^{l}\left\vert x-I\right\vert
_{2}\right) \\
=%
%TCIMACRO{\tsum \limits_{J\in G_{l}^{0}}}%
%BeginExpansion
{\textstyle\sum\limits_{J\in G_{l}^{0}}}
%EndExpansion
\left\{  A_{JI}-\gamma_{I}\delta_{JI}\right\}  2^{\frac{l}{2}}\Omega\left(
2^{l}\left\vert x-J\right\vert _{2}\right)  ,
\end{gather*}
where $\delta_{JI}$ is the Konecker delta. Consequently, operator $m\left.
\boldsymbol{J}_{\mathcal{G}}\right\vert _{\mathcal{X}_{l}}:\mathcal{X}%
_{l}\left(  \mathbb{Z}_{2}\right)  \rightarrow\mathcal{X}_{l}\left(
\mathbb{Z}_{2}\right)  $ is represented by the symmetric matrix
\begin{align*}
mJ_{\mathcal{G}}^{\left(  l\right)  }  &  =m\left[  J_{J,I}^{\left(  l\right)
}\right]  _{J,I\in G_{l}^{0}}=\left[  mA_{JI}-\gamma_{I}m\delta_{JI}\right]
_{J,I\in G_{l}^{0}}\\
& \\
&  =\left\{
\begin{array}
[c]{lll}%
m & \text{if} & J\neq I\text{ and }A_{JI}=1\\
&  & \\
0 & \text{if} & J\neq I\text{ and }A_{JI}=0\\
&  & \\
-m\mathrm{val}(I) & \text{if} & J=I.
\end{array}
\right.
\end{align*}
The matrix of the potential is $V^{(l)}=$\textrm{diag}$\left[  V_{J}\right]
_{J\in G_{l}^{0}}$. We set
\begin{equation}
H^{\left(  l\right)  }=-mJ_{\mathcal{G}}^{\left(  l\right)  }+V^{(l)}=\left[
H_{J,I}^{\left(  l\right)  }\right]  _{J,I\in G_{l}^{0}},
\label{Maytrix_F_G_Eq}%
\end{equation}
where%
\[
H_{J,I}^{\left(  l\right)  }=\left\{
\begin{array}
[c]{lll}%
-m & \text{if} & J\neq I\text{ and }A_{JI}=1\\
&  & \\
0 & \text{if} & J\neq I\text{ and }A_{JI}=0\\
&  & \\
m\mathrm{val}(I)+V_{I} & \text{if} & J=I.
\end{array}
\right.
\]
The discretization of the $2$-adic Schr\"{o}dinger equation takes the form
\begin{equation}
i\frac{\partial}{\partial t}\left[  \Psi_{I}^{\left(  l\right)  }\left(
t\right)  \right]  =H^{\left(  l\right)  }\left[  \Psi_{I}^{\left(  l\right)
}\left(  t\right)  \right]  \text{, }t\geq0. \label{Eq_11}%
\end{equation}

\subsection{The Farhi-Gutmann CTQWs}

Let $\mathcal{G}$ be a finite graph. We take $G_{l}^{0}=V(\mathcal{G})$, the
set of vertices, and%
\[
\Psi\left(  t\right)  :=%
%TCIMACRO{\dsum \limits_{I\in V(\mathcal{G})}}%
%BeginExpansion
{\displaystyle\sum\limits_{I\in V(\mathcal{G})}}
%EndExpansion
\Psi_{I}^{\left(  l\right)  }\left(  t\right)  \left\vert e_{I}\right\rangle =%
%TCIMACRO{\dsum \limits_{I\in V(\mathcal{G})}}%
%BeginExpansion
{\displaystyle\sum\limits_{I\in V(\mathcal{G})}}
%EndExpansion
\left\langle e_{I}\right\vert \left.  \Psi\left(  t\right)  \right\rangle
\left\vert e_{I}\right\rangle ,
\]
with
\[
\left\Vert \Psi\left(  t\right)  \right\Vert ^{2}=%
%TCIMACRO{\dsum \limits_{I\in V(\mathcal{G})}}%
%BeginExpansion
{\displaystyle\sum\limits_{I\in V(\mathcal{G})}}
%EndExpansion
\left\vert \left\langle e_{I}\right\vert \left.  \Psi\left(  t\right)
\right\rangle \right\vert ^{2}=1.
\]
Now, we take $V=0$, and assume that the entries in the diagonal of the
adjacency matrix $\left[  A_{JI}\right]  _{J,I\in G_{N}^{0}}$ are zero, and
set
\begin{equation}
\left\langle e_{I}\right\vert \widehat{H}\left\vert e_{K}\right\rangle
:=H_{I,K}^{\left(  l\right)  }. \label{Matrix_Eq}%
\end{equation}
Then, equation (\ref{Eq_11}) can be rewritten as%
\begin{equation}
i\frac{\partial}{\partial t}\left\langle e_{I}\right\vert \left.  \Psi\left(
t\right)  \right\rangle =%
%TCIMACRO{\dsum \limits_{K\in V(\mathcal{G})}}%
%BeginExpansion
{\displaystyle\sum\limits_{K\in V(\mathcal{G})}}
%EndExpansion
\left\langle e_{I}\right\vert \widehat{H}\left\vert e_{K}\right\rangle
\left\langle e_{K}\right\vert \left.  \Psi\left(  t\right)  \right\rangle ,
\label{Farhi-Gutman_Eq}%
\end{equation}
which is the Schr\"{o}dinger equation for the Farhi-Gutmann CTQW, \cite{Childs
et al}, \cite{Farhi-Gutman}.

\subsection{\label{Subsection _Scaling_Limit}Scaling limit}

Starting with equation (\ref{Farhi-Gutman_Eq}), using (\ref{Matrix_Eq}) and
$\mathbb{C}^{2^{l}}\simeq\mathcal{D}_{l}\left(  \mathbb{Z}_{2}\right)  $, we
recast this equation as equation (\ref{Eq_11}), where matrix $H^{\left(
l\right)  }$ is given in (\ref{Maytrix_F_G_Eq}). The operators $\left.
\boldsymbol{J}_{\mathcal{G}}\right\vert _{\mathcal{X}_{l}}$, respectively
$\left.  \boldsymbol{V}\right\vert _{\mathcal{X}_{l}}$, extend to
$\boldsymbol{J}_{\mathcal{G}},\boldsymbol{V}:L^{2}(\mathcal{K}_{l})\rightarrow
L^{2}(\mathcal{K}_{l})$, and consequently, the equation%
\begin{equation}
i\frac{\partial\Psi(x,t)}{\partial t}=\left(  -m\boldsymbol{J}_{\mathcal{G}%
}+\boldsymbol{V}\right)  \Psi(x,t),x\in\mathcal{K}_{l}\subset\mathbb{Z}%
_{2},t\geq0, \label{Eq_SCh_J}%
\end{equation}
can be interpreted as continuous limit (or scale limit) of equation
(\ref{Farhi-Gutman_Eq}). A precise argument is as follows. We cover the open
compact \ $\mathcal{K}=\mathcal{K}_{l}$ by balls of the form $J+2^{r}%
\mathbb{Z}_{2}$, with $J\in\mathcal{K}_{l}\cap G_{r}$, $r>l$. We now consider
the subspace $\mathcal{X}_{r}\left(  \mathcal{K}_{l}\right)  $ of test
functions of the form%
\[
\varphi^{\left(  r\right)  }\left(  x\right)  =%
%TCIMACRO{\dsum \limits_{J\in\mathcal{K}_{l}\cap G_{r}}}%
%BeginExpansion
{\displaystyle\sum\limits_{J\in\mathcal{K}_{l}\cap G_{r}}}
%EndExpansion
\varphi_{J}2^{\frac{r}{2}}\Omega\left(  2^{r}\left\vert x-I\right\vert
_{2}\right)  \text{, }\varphi_{I}\in\mathbb{C}.
\]
Now, applying the discretization techniques, we obtain a finer discretization
of (\ref{Eq_SCh_J}) of the form
\[
i\frac{\partial}{\partial t}\left[  \Psi_{J}^{\left(  r\right)  }\left(
t\right)  \right]  =H^{\left(  r\right)  }\left[  \Psi_{J}^{\left(  r\right)
}\left(  t\right)  \right]  \text{, for }J\in\mathcal{K}_{l}\cap G_{r}.
\]
Here, it important to say that the matrices\ $H^{\left(  r\right)  }$, $r>l$,
are are block-matrices constructed using copies of $H^{\left(  l\right)  }$;
see \cite{zuniga2020reaction} for a similar calculation. The operators
$-m\boldsymbol{J}_{\mathcal{G}}+\boldsymbol{V}$, $H^{\left(  r\right)  }$ are
self-adjoint, so the functions \ $e^{-i\boldsymbol{H}t}\Psi_{0}$,
$e^{-i\boldsymbol{H}^{\left(  r\right)  }t}\boldsymbol{P}_{r}\left(  \Psi
_{0}\right)  $ are well-defined for any $\Psi_{0}\in L^{2}(\mathcal{K}_{l})$.
The precise meaning of the scale limit is%
\begin{equation}
\lim_{r\rightarrow\infty}\left\Vert e^{-i\boldsymbol{H}t}\Psi_{0}%
-e^{-i\boldsymbol{H}^{\left(  r\right)  }t}\boldsymbol{P}_{r}\left(  \Psi
_{0}\right)  \right\Vert _{2}=0, \label{Limit}%
\end{equation}
for any $\Psi_{0}\in L^{2}(\mathcal{K}_{l})$. Here, we should recall that
$H^{\left(  r\right)  }$ is the matrix representing the restriction of
operator $-m\boldsymbol{J}_{\mathcal{G}}+\boldsymbol{V}$ to subspace
$\mathcal{X}_{r}\left(  \mathcal{K}_{l}\right)  $, so $e^{-i\boldsymbol{H}%
^{\left(  r\right)  }t}\boldsymbol{P}_{r}\left(  \Psi_{0}\right)
=e^{-i\boldsymbol{H}t}\boldsymbol{P}_{r}\left(  \Psi_{0}\right)  $, for $r\geq
l$. Then
\begin{align*}
\lim_{r\rightarrow\infty}\left\Vert e^{-i\boldsymbol{H}t}\Psi_{0}%
-e^{-i\boldsymbol{H}^{\left(  r\right)  }t}\boldsymbol{P}_{r}\left(  \Psi
_{0}\right)  \right\Vert _{2}  &  =\lim_{r\rightarrow\infty}\left\Vert
e^{-i\boldsymbol{H}t}\Psi_{0}-e^{-i\boldsymbol{H}t}\boldsymbol{P}_{r}\left(
\Psi_{0}\right)  \right\Vert _{2}\\
\lim_{r\rightarrow\infty}\left\Vert e^{-i\boldsymbol{H}t}\Psi_{0}%
-e^{-i\boldsymbol{H}t}\boldsymbol{P}_{r}\left(  \Psi_{0}\right)  \right\Vert
_{2}  &  \leq\lim_{r\rightarrow\infty}\left\Vert \Psi_{0}-\boldsymbol{P}%
_{r}\left(  \Psi_{0}\right)  \right\Vert _{2}=0.
\end{align*}
The last limit is a reformulation of the fact that $\mathcal{D}(\mathbb{Z}%
_{2})=$ $\cup_{r}\mathcal{D}_{r}(\mathbb{Z}_{2})$, with $\mathcal{D}%
_{r}(\mathbb{Z}_{2})\subset\mathcal{D}_{r+1}(\mathbb{Z}_{2})$, is dense in
$L^{2}(\mathbb{Z}_{2})$, since $L^{2}\left(  \mathcal{K}_{l}\right)  \subset
L^{2}(\mathbb{Z}_{2})$, then $\cup_{r}\left(  \mathcal{K}\cap\mathcal{D}%
_{r}(\mathbb{Z}_{2})\right)  $ is also dense in the last space.

The above can be extended to the Schr\"{o}dinger equations of type
(\ref{Eq_SChrodinger_4}). So, we can say that the $2$-adic Schr\"{o}dinger
equations introduced here describe the scaling limit of a CTQWs on undirected
graphs; these CTQWs are generalizations of the ones given in
\cite{Farhi-Gutman}, \cite{Mulkne-Blumen}.

We conjecture that any $p$-adic Schr\"{o}dinger equation, obtained by a Wick
rotation from a $p$-adic \ heat equation, describes \ the scaling limit of a
CTQW on a graph.

\section{\label{Sect_6}CTQWs on bi-weighted graphs}

We now give a further generalization of the Farhi-Gutmann CTQW to a class of
oriented graphs, that we have called bi-weighted graphs. We take $\mathcal{G}$
a graph, with vertices $G_{l}^{0}\subset G_{l}$ as before. We endow
$\mathcal{G}$ with two weighted symmetric matrices $A=\left[  A_{K,I}\right]
_{K,I\in G_{l}^{0}}$, $B=\left[  B_{K,I}\right]  _{K,I\in G_{l}^{0}}$, with
non-negative entries. The entry $A_{K,I}$ is interpreted as an outward flux
rate from node $I$ to node $K$; \ while the entry $B_{K,I}$ is interpreted as
an inward flux rate from $K$ to $I$. The discretization of
(\ref{Eq_SChrodinger_4}) is the Schr\"{o}dinger equation on a bi-weighted graph.

We assume that the approximations of the functions $A(x,y)$, $B(x,y)$ are%
\begin{equation}
A^{\left(  l\right)  }(x,y)=%
%TCIMACRO{\dsum \limits_{I\in G_{l}^{0}}}%
%BeginExpansion
{\displaystyle\sum\limits_{I\in G_{l}^{0}}}
%EndExpansion%
%TCIMACRO{\dsum \limits_{K\in G_{l}^{0}}}%
%BeginExpansion
{\displaystyle\sum\limits_{K\in G_{l}^{0}}}
%EndExpansion
2^{l}A_{I,K}\Omega\left(  2^{l}\left\vert x-I\right\vert _{2}\right)
\Omega\left(  2^{l}\left\vert y-K\right\vert _{2}\right)  , \label{Eq_A}%
\end{equation}%
\begin{equation}
B^{\left(  l\right)  }(x,y)=%
%TCIMACRO{\dsum \limits_{I\in G_{l}^{0}}}%
%BeginExpansion
{\displaystyle\sum\limits_{I\in G_{l}^{0}}}
%EndExpansion%
%TCIMACRO{\dsum \limits_{K\in G_{l}^{0}}}%
%BeginExpansion
{\displaystyle\sum\limits_{K\in G_{l}^{0}}}
%EndExpansion
2^{l}B_{I,K}\Omega\left(  2^{l}\left\vert x-I\right\vert _{2}\right)
\Omega\left(  2^{l}\left\vert y-K\right\vert _{2}\right)  . \label{Eq_B}%
\end{equation}

We set%
\[
\gamma_{I}^{\left(  A\right)  }:=%
%TCIMACRO{\dsum \limits_{K\in G_{l}^{0}}}%
%BeginExpansion
{\displaystyle\sum\limits_{K\in G_{l}^{0}}}
%EndExpansion
A_{I,K},\text{ \ }\gamma_{I}^{\left(  B\right)  }:=%
%TCIMACRO{\dsum \limits_{K\in G_{l}^{0}}}%
%BeginExpansion
{\displaystyle\sum\limits_{K\in G_{l}^{0}}}
%EndExpansion
B_{I,K}.
\]

We approximate $\Psi\left(  x,t\right)  $ as
\[
\Psi^{\left(  l\right)  }\left(  x,t\right)  =%
%TCIMACRO{\dsum \limits_{I\in G_{l}^{0}}}%
%BeginExpansion
{\displaystyle\sum\limits_{I\in G_{l}^{0}}}
%EndExpansion
2^{\frac{l}{2}}\Psi_{I}^{\left(  l\right)  }\left(  t\right)  \Omega\left(
2^{l}\left\vert x-I\right\vert _{2}\right)  .
\]
Then%
\begin{align}
V^{\left(  l\right)  }(x)  &  =%
%TCIMACRO{\dsum \limits_{I\in G_{l}^{0}}}%
%BeginExpansion
{\displaystyle\sum\limits_{I\in G_{l}^{0}}}
%EndExpansion
\left\{
%TCIMACRO{\dsum \limits_{K\in G_{l}^{0}}}%
%BeginExpansion
{\displaystyle\sum\limits_{K\in G_{l}^{0}}}
%EndExpansion
\left(  B_{I,K}-A_{I,K}\right)  \right\}  \Omega\left(  2^{l}\left\vert
x-I\right\vert _{2}\right) \label{Eq_V0}\\
&  =%
%TCIMACRO{\dsum \limits_{I\in G_{l}^{0}}}%
%BeginExpansion
{\displaystyle\sum\limits_{I\in G_{l}^{0}}}
%EndExpansion
\left\{  \gamma_{I}^{\left(  B\right)  }-\gamma_{I}^{\left(  A\right)
}\right\}  \Omega\left(  2^{l}\left\vert x-I\right\vert _{2}\right)
,\nonumber
\end{align}
see (\ref{Potential}), and%
\[
\Psi^{\left(  l\right)  }\left(  x,t\right)  V^{\left(  l\right)  }(x)=%
%TCIMACRO{\dsum \limits_{I\in G_{l}^{0}}}%
%BeginExpansion
{\displaystyle\sum\limits_{I\in G_{l}^{0}}}
%EndExpansion
\left\{  \gamma_{I}^{\left(  B\right)  }-\gamma_{I}^{\left(  A\right)
}\right\}  \Psi_{I}^{\left(  l\right)  }\left(  t\right)  2^{\frac{l}{2}%
}\Omega\left(  2^{l}\left\vert x-I\right\vert _{2}\right)  .
\]
Now
\begin{gather*}
\int\limits_{\mathcal{K}_{l}}A^{\left(  l\right)  }(x,y)\left\{  \Psi^{\left(
l\right)  }(y,t)-\Psi^{\left(  l\right)  }(x,t)\right\}  dy=\\%
%TCIMACRO{\dsum \limits_{I\in G_{l}^{0}}}%
%BeginExpansion
{\displaystyle\sum\limits_{I\in G_{l}^{0}}}
%EndExpansion%
%TCIMACRO{\dsum \limits_{K\in G_{l}^{0}}}%
%BeginExpansion
{\displaystyle\sum\limits_{K\in G_{l}^{0}}}
%EndExpansion
2^{l}A_{I,K}\Omega\left(  2^{l}\left\vert x-I\right\vert _{2}\right)
\int\limits_{\mathcal{K}_{l}}\Omega\left(  2^{l}\left\vert y-K\right\vert
_{2}\right)  \left\{  \Psi^{\left(  l\right)  }(y,t)-\Psi^{\left(  l\right)
}(x,t)\right\}  dy\\
=%
%TCIMACRO{\dsum \limits_{I\in G_{l}^{0}}}%
%BeginExpansion
{\displaystyle\sum\limits_{I\in G_{l}^{0}}}
%EndExpansion%
%TCIMACRO{\dsum \limits_{K\in G_{l}^{0}}}%
%BeginExpansion
{\displaystyle\sum\limits_{K\in G_{l}^{0}}}
%EndExpansion
2^{l}A_{I,K}\Omega\left(  2^{l}\left\vert x-I\right\vert _{2}\right)  \left\{
2^{\frac{-l}{2}}\Psi_{K}^{\left(  l\right)  }(t)-2^{-l}\Psi^{\left(  l\right)
}(x,t)\right\} \\
=%
%TCIMACRO{\dsum \limits_{I\in G_{l}^{0}}}%
%BeginExpansion
{\displaystyle\sum\limits_{I\in G_{l}^{0}}}
%EndExpansion%
%TCIMACRO{\dsum \limits_{K\in G_{l}^{0}}}%
%BeginExpansion
{\displaystyle\sum\limits_{K\in G_{l}^{0}}}
%EndExpansion
A_{I,K}\Psi_{K}^{\left(  l\right)  }(t)2^{\frac{l}{2}}\Omega\left(
2^{l}\left\vert x-I\right\vert _{2}\right)  -%
%TCIMACRO{\dsum \limits_{I\in G_{l}^{0}}}%
%BeginExpansion
{\displaystyle\sum\limits_{I\in G_{l}^{0}}}
%EndExpansion%
%TCIMACRO{\dsum \limits_{K\in G_{l}^{0}}}%
%BeginExpansion
{\displaystyle\sum\limits_{K\in G_{l}^{0}}}
%EndExpansion
A_{I,K}\Psi_{I}^{\left(  l\right)  }(t)2^{\frac{l}{2}}\Omega\left(
2^{l}\left\vert x-I\right\vert _{2}\right) \\
=%
%TCIMACRO{\dsum \limits_{I\in G_{l}^{0}}}%
%BeginExpansion
{\displaystyle\sum\limits_{I\in G_{l}^{0}}}
%EndExpansion%
%TCIMACRO{\dsum \limits_{K\in G_{l}^{0}}}%
%BeginExpansion
{\displaystyle\sum\limits_{K\in G_{l}^{0}}}
%EndExpansion
A_{I,K}\Psi_{K}^{\left(  l\right)  }(t)2^{\frac{l}{2}}\Omega\left(
2^{l}\left\vert x-I\right\vert _{2}\right)  -%
%TCIMACRO{\dsum \limits_{I\in G_{l}^{0}}}%
%BeginExpansion
{\displaystyle\sum\limits_{I\in G_{l}^{0}}}
%EndExpansion
\gamma_{I}^{\left(  A\right)  }\Psi_{I}^{\left(  l\right)  }(t)2^{\frac{l}{2}%
}\Omega\left(  2^{l}\left\vert x-I\right\vert _{2}\right) \\
=%
%TCIMACRO{\dsum \limits_{I\in G_{l}^{0}}}%
%BeginExpansion
{\displaystyle\sum\limits_{I\in G_{l}^{0}}}
%EndExpansion
\left\{
%TCIMACRO{\dsum \limits_{K\in G_{l}^{0}}}%
%BeginExpansion
{\displaystyle\sum\limits_{K\in G_{l}^{0}}}
%EndExpansion
A_{I,K}\Psi_{K}^{\left(  l\right)  }(t)-\gamma_{I}^{\left(  A\right)  }%
\Psi_{I}^{\left(  l\right)  }(t)\right\}  2^{\frac{l}{2}}\Omega\left(
2^{l}\left\vert x-I\right\vert _{2}\right) \\
=%
%TCIMACRO{\dsum \limits_{I\in G_{l}^{0}}}%
%BeginExpansion
{\displaystyle\sum\limits_{I\in G_{l}^{0}}}
%EndExpansion
\left\{
%TCIMACRO{\dsum \limits_{K\in G_{l}^{0}}}%
%BeginExpansion
{\displaystyle\sum\limits_{K\in G_{l}^{0}}}
%EndExpansion
\left(  A_{I,K}-\gamma_{I}^{\left(  A\right)  }\delta_{I,K}\right)  \Psi
_{K}^{\left(  l\right)  }(t)\right\}  2^{\frac{l}{2}}\Omega\left(
2^{l}\left\vert x-I\right\vert _{2}\right)  .
\end{gather*}
Therefore, the discretization of (\ref{Eq_SChrodinger_4}) is
\begin{align*}
i\frac{\partial}{\partial t}\left[  \Psi_{I}^{\left(  l\right)  }\left(
t\right)  \right]   &  =-\left(  \left[  A_{K,I}\right]  -\text{\textrm{diag}%
}\left(  \gamma_{I}^{\left(  A\right)  }\right)  \right)  \left[  \Psi
_{K}^{\left(  l\right)  }\left(  t\right)  \right]  +\text{\textrm{diag}%
}\left(  \gamma_{I}^{\left(  B\right)  }-\gamma_{I}^{\left(  A\right)
}\right)  \left[  \Psi_{I}^{\left(  l\right)  }\left(  t\right)  \right] \\
&  =\left(  -\left[  A_{K,I}\right]  +\text{\textrm{diag}}\left(  \gamma
_{I}^{\left(  B\right)  }\right)  \right)  \left[  \Psi_{K}^{\left(  l\right)
}\left(  t\right)  \right]  _{K},
\end{align*}
i.e.,
\[
i\frac{\partial}{\partial t}\left[  \Psi_{I}^{\left(  l\right)  }\left(
t\right)  \right]  =\left(  -A+\text{\textrm{diag}}\left(  \gamma_{I}^{\left(
B\right)  }\right)  \right)  \left[  \Psi_{I}^{\left(  l\right)  }\left(
t\right)  \right]  =-\boldsymbol{H}^{\left(  l\right)  }\left[  \Psi
_{I}^{\left(  l\right)  }\left(  t\right)  \right]  \text{.}%
\]

\section{\label{Appendix} Appendix: basic facts on $2$-adic analysis}

In this section, we fix the notation and collect some basic results on
$2$-adic analysis that we use in this article. For a detailed exposition on
$p$-adic analysis, the reader may consult \cite{V-V-Z}, \cite{Zuniga-Textbook}%
, \cite{Taibleson}, \cite{Alberio et al}.

\subsection{The field of $2$-adic numbers}

The field of $2-$adic numbers $\mathbb{Q}_{2}$ is defined as the completion of
the field of rational numbers $\mathbb{Q}$ with respect to the $2-$adic norm
$|\cdot|_{2}$, which is defined as
\[
|x|_{2}=%
\begin{cases}
0 & \text{if }x=0\\
2^{-\gamma} & \text{if }x=2^{\gamma}\dfrac{a}{b},
\end{cases}
\]
where $a$ and $b$ are integers coprime with $2$. The integer $\gamma
=ord_{2}(x):=ord(x)$, with $ord(0):=+\infty$, is called the\textit{\ }$2-$adic
order of $x$.

The metric space $\left(  \mathbb{Q}_{2},|\cdot|_{2}\right)  $ is a complete
ultrametric space. As a topological space $\mathbb{Q}_{2}$\ is homeomorphic to
a Cantor-like subset of the real line, see, e.g., \cite{V-V-Z}, \cite{Alberio
et al}.

Any $2-$adic number $x\neq0$ has a unique expansion of the form
\[
x=2^{ord(x)}\sum_{j=0}^{\infty}x_{j}2^{j},
\]
where $x_{j}\in\{0,1\}$ and $x_{0}\neq0$. \ In addition, any $x\in
\mathbb{Q}_{2}\smallsetminus\left\{  0\right\}  $ can be represented uniquely
as $x=2^{ord(x)}v$, where $\left\vert v\right\vert _{2}=1$.

\subsection{Topology of $\mathbb{Q}_{2}$}

For $r\in\mathbb{Z}$, denote by $B_{r}(a)=\{x\in\mathbb{Q}_{2};|x-a|_{2}%
\leq2^{r}\}$ the ball of radius $2^{r}$ with center at $a\in\mathbb{Q}_{2}$,
and take $B_{r}(0):=B_{r}$. The ball $B_{0}$ equals $\mathbb{Z}_{2}$, the ring
of $2-$adic integers. The balls are both open and closed subsets in
$\mathbb{Q}_{2}$. In addition, two balls in $\mathbb{Q}_{2}$ are either
disjoint or one is contained in the other. As a topological space $\left(
\mathbb{Q}_{2},|\cdot|_{2}\right)  $ is totally disconnected, i.e., the only
connected \ subsets of $\mathbb{Q}_{2}$ are the empty set and the points. A
subset of $\mathbb{Q}_{2}$ is compact if and only if it is closed and bounded
in $\mathbb{Q}_{2}$, see, e.g., \cite[Section 1.3]{V-V-Z}, or \cite[Section
1.8]{Alberio et al}. The balls are compact subsets. Thus $\left(
\mathbb{Q}_{2},|\cdot|_{2}\right)  $ is a locally compact topological space.

\subsection{The Haar measure}

Since $(\mathbb{Q}_{2},+)$ is a locally compact topological group, there
exists a Haar measure $d^{{}}x$, which is invariant under translations, i.e.,
$d(x+a)=dx$, \cite{Halmos}. If we normalize this measure by the condition
$\int_{\mathbb{Z}_{2}}dx=1$, then $dx$ is unique.

\begin{notation}
We will use $\Omega\left(  2^{-r}|x-a|_{2}\right)  $ to denote the
characteristic function of the ball $B_{r}(a)=a+2^{-r}\mathbb{Z}_{2}$, where
\[
\mathbb{Z}_{2}=\left\{  x\in\mathbb{Q}_{2};\left\vert x\right\vert _{2}%
\leq1\right\}
\]
is the unit ball. For more general sets, we will use the notation $1_{A}$ for
the characteristic function of set $A$.
\end{notation}

\subsection{The Bruhat-Schwartz space}

A complex-valued function $\varphi$ defined on $\mathbb{Q}_{2}$ is called
locally constant if for any $x\in\mathbb{Q}_{2}$ there exist an integer
$l(x)\in\mathbb{Z}$ such that%
\begin{equation}
\varphi(x+x^{\prime})=\varphi(x)\text{ for any }x^{\prime}\in B_{l(x)}.
\label{local_constancy}%
\end{equation}
A function $\varphi:\mathbb{Q}_{2}\rightarrow\mathbb{C}$ is called a
Bruhat-Schwartz function (or a test function) if it is locally constant with
compact support. Any test function can be represented as a linear combination,
with complex coefficients, of characteristic functions of balls. The
$\mathbb{C}$-vector space of Bruhat-Schwartz functions is denoted by
$\mathcal{D}(\mathbb{Q}_{2})$. For $\varphi\in\mathcal{D}(\mathbb{Q}_{2})$,
the largest number $l=l(\varphi)$ satisfying (\ref{local_constancy}) is called
the exponent of local constancy (or the parameter of constancy) of $\varphi$.

\subsection{$L^{\rho}$ spaces}

Given $\rho\in\lbrack1,\infty)$, we denote by$L^{\rho}\left(
%TCIMACRO{\U{211a} }%
%BeginExpansion
\mathbb{Q}
%EndExpansion
_{2}\right)  :=L^{\rho}\left(
%TCIMACRO{\U{211a} }%
%BeginExpansion
\mathbb{Q}
%EndExpansion
_{2},dx\right)  ,$ the $\mathbb{C}-$vector space of all the complex valued
functions $g$ satisfying
\[
\left\Vert g\right\Vert _{\rho}=\left(  \text{ }%
%TCIMACRO{\dint \limits_{\mathbb{Q}_{2}}}%
%BeginExpansion
{\displaystyle\int\limits_{\mathbb{Q}_{2}}}
%EndExpansion
\left\vert g\left(  x\right)  \right\vert ^{\rho}dx\right)  ^{\frac{1}{\rho}%
}<\infty,
\]
where $dx$ is the normalized Haar measure on $\left(  \mathbb{Q}_{2},+\right)
$.

If $U$ is an open subset of $\mathbb{Q}_{2}$, $\mathcal{D}(U)$ denotes the
$\mathbb{C}$-vector space of test functions with supports contained in $U$,
then $\mathcal{D}(U)$ is dense in
\[
L^{\rho}\left(  U\right)  =\left\{  \varphi:U\rightarrow\mathbb{C};\left\Vert
\varphi\right\Vert _{\rho}=\left\{
%TCIMACRO{\dint \limits_{U}}%
%BeginExpansion
{\displaystyle\int\limits_{U}}
%EndExpansion
\left\vert \varphi\left(  x\right)  \right\vert ^{\rho}dx\right\}  ^{\frac
{1}{\rho}}<\infty\right\}  ,
\]
for $1\leq\rho<\infty$, see, e.g., \cite[Section 4.3]{Alberio et al}.

\bigskip

\subsection{Acknowledgements}

The author has no funding source to declare.

\subsection{Conflict of Interest}

The author declares no conflict of interest.

\subsection{Data Availability and Declaration of Interests}

No data were utilized in this theoretical paper.

\subsection{Keywords}

$2$- adic quantum mechanics, $p$-adic numbers, continuous-time random walk on
graphs, Planck length.

\end{document}